\documentclass[11pt,onecolumn,draftcls,journal]{IEEEtran}
\usepackage{amsmath,amssymb,amsthm, graphicx,float,cite,bm}
\usepackage{algorithm,algorithmic,multirow,MnSymbol}
\usepackage{setspace,hyperref,array,subfig,relsize}
\usepackage[usenames,dvipsnames]{color}

\hypersetup{
    colorlinks=false,
    pdfborder={0 0 0},
}

\allowdisplaybreaks[4]

\newtheorem{thm}{Theorem}
\newtheorem{lem}{Lemma}
\newtheorem{defn}{Definition}

\newtheorem{rmk}{Remark}
\newtheorem{asmp}{Assumption}
\newtheorem{fct}{Fact}
\newtheorem{cor}{Corollary}
\newcolumntype{C}[1]{>{\centering\let\newline\\\arraybackslash\hspace{0pt}}m{#1}}

\DeclareMathOperator{\diag}{diag}
\DeclareMathOperator{\tr}{tr}

\DeclareMathAlphabet\mathbfcal{OMS}{cmsy}{b}{n}
\providecommand{\norm}[1]{\lVert#1\rVert}
\DeclareMathOperator{\blkdiag}{blkdiag}

\begin{document}

\title{\LARGE \bf Nonlinear POMDPs for Active State Tracking \\ with Sensing Costs}
\author{Daphney-Stavroula Zois,$^\star$ \textit{Student Member, IEEE}, and Urbashi Mitra, \textit{Fellow, IEEE}
\thanks{D.-S. Zois and U. Mitra are with the Ming Hsieh Department of Electrical Engineering, University of Southern California, Los Angeles, CA. e-mail: $\lbrace$zois, ubli$\rbrace$ @usc.edu. 3740 McClintock Ave., Los Angeles, CA 90089-2565.}
\thanks{This research has been funded in part by the following grants and organizations: ONR N00014-09-1-0700, NSF CNS-0832186,
CCF-0917343, CCF-1117896, CNS-1213128,  AFOSR FA9550-12-1-0215, DOT CA-26-7084-00, the National Center on Minority Health and Health Disparities (NCMHD) (supplement to P60 MD002254), Nokia and Qualcomm.}
\thanks{Parts of the material in this paper have been previously presented at Asilomar 2013 and GlobalSIP 2013.}}

\maketitle

\begin{abstract}

Active state tracking is needed in object classification, target tracking, medical diagnosis and estimation of sparse signals among other various applications. Herein, active state tracking of a discrete--time, finite--state Markov chain is considered. Noisy Gaussian observations are dynamically collected by exerting appropriate control over their information content, while incurring a related sensing cost. The objective is to devise sensing strategies to optimize the trade--off between tracking performance and sensing cost. A recently proposed Kalman--like estimator \cite{ZoisTSP14} is employed for state tracking. The associated mean--squared error and a generic sensing cost metric are then used in a partially observable Markov decision process formulation, and the optimal sensing strategy is derived via a dynamic programming recursion. The resulting recursion proves to be non--linear, challenging control policy design. Properties of the related cost functions are derived and sufficient conditions are provided regarding the structure of the optimal control policy enabling characterization of when passive state tracking is optimal. To overcome the associated computational burden of the optimal sensing strategy, two lower complexity strategies are proposed, which exploit the aforementioned properties. The performance of the proposed strategies is illustrated in a wireless body sensing application, where cost savings as high as $60\%$ are demonstrated for a $4\%$ detection error with respect to a static equal allocation sensing strategy. 
\end{abstract}

\section{Introduction}

Active state tracking is a generalization of the classical state tracking problem. In particular, the objective is to \textit{accurately} and \textit{efficiently} track the unknown state of a dynamical system by \textit{adaptively} exploiting different sensing capabilities (\emph{e.g.} sensor type, number of samples, location) as a function of past information. In contrast to traditional control systems, where control affects system state evolution, in active state tracking applications, the controller actively selects between the available observations, but does not affect the plant. Applications include: object classification, target tracking \cite{AtiaTSP11}, context awareness \cite{WangIPSN10}, health care \cite{ZoisTSP13}, estimation of sparse signals \cite{WeiJSTSP13}, and coding with feedback \cite{NaghshvararXiv13}.

In this paper, we study the active state tracking problem for systems modeled by discrete--time, finite--state Markov chains. We dynamically select between noisy Gaussian measurement vectors by exerting appropriate control over their information content, while incurring a sensing cost. Our goal is to devise sensing strategies to optimize the trade--off between tracking performance and sensing cost. To this end, we propose a partially observable Markov decision process (POMDP) formulation and adopt our earlier proposed approximate minimum mean--squared error (MMSE) estimator \cite{ZoisTSP14} for state tracking.

Our current and previous work \cite{ZoisTSP14} differ as follows:
\begin{enumerate}
\item[i.] Herein, our goal is to optimize the trade--off between tracking performance and sensing cost versus our prior work, where we only optimized tracking performance.
\item[ii.] We derive the optimal sensing strategy for this new optimization problem via dynamic programming (DP). In contrast to \cite{ZoisTSP14}, we also derive properties of the cost--to--go function and sufficient conditions for the structure of the optimal sensing strategy.
\item[iii.] Finally, we propose two lower complexity sensing strategies to circumvent the high computational complexity associated with the optimal sensing strategy in contrast to \cite{ZoisTSP14}, where only the optimal sensing strategy for optimizing tracking performance was considered.
\end{enumerate}

In recent years, active state tracking has received considerable research attention. Both the static \cite{NaghshvarAS13, NitinawaratTAC13}, \emph{i.e.} the system state does \underline{not} change with time, and the time--varying \cite{KrishnamurthyTSP07, HernandezTAES04, AtiaTSP11, MasazadeCISS12, ZoisTSP13, KrishnamurthyIT13} case have been previously considered. For the latter, most prior work assumes discrete observations \cite{KrishnamurthyTSP07, MasazadeCISS12, ZoisTSP13, KrishnamurthyIT13}, scalar \cite{AtiaTSP11, KrishnamurthyIT13} or $w$ independent measurements from $w$ sensors \cite{AtiaTSP11}. In contrast, we focus on time--varying systems with Gaussian measurement vectors, which also account for fusion of multiple different types of measurements.

A variety of cost functions has been previously adopted as performance quality measures, such as detection error probability and bounds \cite{KrishnamurthyTSP07, ZoisTSP13, KrishnamurthyIT13, NitinawaratTAC13, NaghshvarAS13}, mean--squared error (MSE) \cite{KrishnamurthyTSP07, WeiJSTSP13, ZoisTSP14}, information--theoretic measures \cite{KrishnamurthyTSP07}, distance metrics \cite{AtiaTSP11} and estimation bounds \cite{HernandezTAES04, MasazadeCISS12}. Similar to \cite{KrishnamurthyTSP07, WeiJSTSP13, ZoisTSP14}, we focus on MSE because 1) we wish to optimize the \emph{belief state}, which is the MMSE state estimate and constitutes a very good indicator of the unknown system state, and 2) we can acquire closed--form formulae for the MSE performance, which enable us to explicitly focus on true estimation performance versus other metrics, which do not admit closed form solutions, and their approximation can affect the sensing strategy. Contrary to \cite{HernandezTAES04, MasazadeCISS12, NitinawaratTAC13, NaghshvarAS13, ZoisTSP14}, we adopt sensing usage costs. In most cases, the associated POMDP is linear \cite{AtiaTSP11, ZoisTSP13, NaghshvarAS13, KrishnamurthyIT13, NitinawaratTAC13} in the belief state resulting in a standard formulation that is in general easier to characterize since the relevant value function is known to be piecewise linear and convex \cite{LovejoyAOR91}. In contrast, our POMDP is \underline{non--linear} and thus, harder to characterize. Non--linear POMDPs have previously appeared in \cite{KrishnamurthyTSP07}, where only one out a finite number of sensors can be selected and the MSE metric employed by the authors is scaled by a user--defined cost in an effort to capture the effect of different sensors. In contrast, our framework is more widely applicable since it allows the selection of multiple heterogeneous sensors, while their effect is directly captured by our MSE metric without the need of additional user--defined variables.

Sufficient conditions under which active sensing reduces to passive sensing and the optimal sensing policy has a threshold structure for linear and non--linear POMDPs have been previously derived in \cite{NaghshvarISIT10} and \cite{KrishnamurthyTSP07}, respectively. In contrast to \cite{NaghshvarISIT10}, for the two--state case with scalar measurements, we establish the concavity of the cost--to--go function for our non--linear POMDP and generalize the conditions of \cite{NaghshvarISIT10} in three ways: we consider 1) non--linear POMDPs, 2) time--varying system states, and 3) different sensing usage costs. We also illustrate cases where active sensing is unavoidable and provide the exact form of the threshold. Note that we do not impose any restrictive constraints on the effect of controls on the belief state evolution versus \cite{KrishnamurthyTSP07}, where a ``quantized" evolution is imposed. A broad spectrum of applications can be formulated as a two--state problem with scalar measurements, \emph{e.g.} spectrum sensing for cognitive radio \cite{UnnikrishnanTSP10}, collision prediction for intelligent transportation \cite{AtevTITS05}, user motion estimation for context awareness \cite{WangIPSN10}, and outlier detection \cite{HodgeAIR04}.

Dynamic programming is prohibitive for large problem sizes. We propose two lower complexity sensing strategies with efficient implementations: a myopic strategy and a strategy, where the Weiss--Weinstein lower bound (WWLB) \cite{WeinsteinIT88} is used instead of the MSE. The WWLB provides a theoretical performance limit for a Bayesian estimator, and is essentially free from \emph{regularity conditions}\footnote{The regularity conditions refer to the existence of derivatives of the joint pdf of the observations and the parameters.}, versus other well--known bounds, \emph{e.g.} the Cram\'er--Rao lower bound (CRLB), the Bhattacharyya lower bound (BLB) \cite{VanTrees07}, and thus, it is applicable to the estimation of discrete parameters. Contrary to sensor selection algorithms based on the Bayesian CRLB \cite{MasazadeCISS12, HernandezTAES04}, we optimize the trade--off between the WWLB and sensing cost. We derive closed--form formulae for the sequential WWLB \cite{ReeceFusion05, RapoportTSP07, XaverTSP13} for our system model, accounting for discrete parameters and control inputs versus \cite{HernandezTAES04}, where numerical methods were employed to approximate key terms, and \cite{MasazadeCISS12}, where key posterior distributions were approximated. Prior work on sequential WWLBs has focused on continuous parameters \cite{ReeceFusion05}, discretized versions of continuous parameters \cite{XaverTSP13} or two--valued discrete parameters with restrictive assumptions on the bound \cite{RapoportTSP07}, without exerting control. To the best of our knowledge, we are the first to design a sensing strategy based on the optimization of WWLB for multi--valued discrete parameters.

Our contributions are as follows. For the active state tracking problem, we propose a POMDP formulation to optimize the trade--off between MSE and a sensing cost metric, and derive the optimal sensing strategy using DP. For the case of two states and scalar measurements, we establish the concavity of the cost--to--go function and give sufficient conditions under which passive sensing is optimal. We also illustrate how decision making is accomplished (\emph{cf.} threshold structure) when active sensing is required. Even though DP constitutes the standard way of determining the optimal sensing strategy, the \emph{curse of dimensionality} (\emph{i.e.} one or all of the state, observation and control spaces are large) makes it impractical for large--scale applications. Furthermore, the nonlinear structure of our POMDP further challenges control policy determination. To overcome the associated computational burden, we propose a myopic strategy\footnote{Due the concavity of the cost--to--go function, the associated strategy has a very nice structure, known as \emph{threshold structure}, for the special case of two states and scalar measurements.}, and a cost--efficient WWLB (CE--WWLB) strategy. For the latter, we first derive closed--form expressions for the sequential WWLB in the case of multi--valued discrete parameters and control inputs. We make connections between the bound and detection performance (\emph{i.e.} the Bhattacharyya coefficient and the Chernoff bound \cite{Duda01}). We validate the performance of the proposed sensing strategies on real data from a body sensing application and observe cost savings as high as $60\%$ with acceptable detection error.

The rest of the paper is organized as follows. In Section \ref{sec:PD}, we present the system model and the optimization problem. We also review our Kalman--like estimator. In Sections \ref{sec:OSS} and \ref{sec:MR}, we give the DP recursion and prove properties of the cost--to--go function and sufficient conditions for the optimal control policy structure, respectively. In Section \ref{sec:LCS}, we propose two lower complexity strategies, and in Section \ref{sec:NR}, we illustrate the performance of the proposed strategies in a body sensing application. We conclude the paper in Section \ref{sec:C}.

\textbf{Notation}. Unless stated, all vectors are column vectors denoted by lowercase boldface symbols (\emph{e.g.} $\mathbf{v}$) and all matrices are denoted by uppercase boldface symbols (\emph{e.g.} $\mathbf{A}$). Sets are denoted by calligraphic symbols (\emph{e.g.} $\mathcal{X}$) and $|\mathcal{X}|$ denotes the cardinality of set $\mathcal{X}$. $\mathbf{1}$ denotes a vector with all components equal to one and $\mathbf{I}$ the identity matrix. $\tr(\cdot)$ denotes the trace operator, $|\mathbf{A}|$ the determinant of matrix $\mathbf{A}$, $\norm{\mathbf{x}}$ the $L^{2}$--norm of vector $\mathbf{x}$, $\diag(\mathbf{x})$ the diagonal matrix with elements the components of vector $\mathbf{x}$ and $\blkdiag(\mathbf{A}_{1},\dotsc,\mathbf{A}_{n})$ the block diagonal matrix with main diagonal blocks the matrices $\mathbf{A}_{1}, \dotsc, \mathbf{A}_{n}$. Finally, for any event $B$, $\mathbf{1}_{B}$ is the indicator function, \emph{i.e.} $\mathbf{1}_{B} = 1$ when $B$ occurs, otherwise $\mathbf{1}_{B} = 0$.

\section{Problem Definition}\label{sec:PD}

In this section, we introduce our formulation and review our previously proposed Kalman--like estimator \cite{ZoisTSP14}.

\subsection{System Model}\label{subsec:SM}

We consider a particular class of dynamical systems known as POMDPs \cite{BertsekasDPOC05}, where time is divided into discrete time slots represented by $k \in \lbrace 0, 1, \dotsc \rbrace$. The system state at time slot $k$, denoted by $\mathbf{x}_{k}$, is modeled by a finite--state, first--order Markov chain with $n = |\mathcal{X}|$ states, where $\mathcal{X} = \lbrace \mathbf{e}_{1}, \mathbf{e}_{2}, \dotsc, \mathbf{e}_{n}\rbrace$ and $\mathbf{e}_{i}$ represents a $n$--dimensional unit vector with one in the $i$th position and zeros everywhere else. The Markov chain statistics are described by a $n \times n$ transition probability matrix $\mathbf{P}$ with elements $P_{j|i} = P(\mathbf{x}_{k+1} = \mathbf{e}_{j} | \mathbf{x}_{k} = \mathbf{e}_{i})$, $\forall \mathbf{e}_{i}, \mathbf{e}_{j} \in \mathcal{X}$. We assume that the Markov chain is stationary, \emph{i.e.} the related state transition probabilities do not change with time.

At each time slot, the exact value of the current state is unknown. Instead, the controller decides to receive all or a subset of noisy observations by selecting the appropriate control input $\mathbf{u}_{k-1}$ at the end of time slot $k-1$. Thus, at time slot $k$, a measurement vector $\mathbf{y}_{k}$ is received, which is described by the multivariate Gaussian observation kernel of the form
\begin{equation}\label{eq:observation_model}
\mathbf{y}_{k} \big | \mathbf{e}_{i}, \mathbf{u}_{k-1} \sim f(\mathbf{y}_{k}|\mathbf{e}_{i},\mathbf{u}_{k-1}) = \mathcal{N}(\mathbf{m}_{i}^{\mathbf{u}_{k-1}}, \mathbf{Q}_{i}^{\mathbf{u}_{k-1}} )
\end{equation}

\noindent
for all $\mathbf{e}_{i} \in \mathcal{X}$. We denote by $\mathbf{m}_{i}^{\mathbf{u}_{k-1}}$ and $\mathbf{Q}_{i}^{\mathbf{u}_{k-1}}$ the conditional mean vector and covariance matrix of the measurement vector for system state $\mathbf{e}_{i}$ and control input $\mathbf{u}_{k-1}$, respectively. We denote by $X^{k} = \lbrace \mathbf{x}_{0}, \mathbf{x}_{1}, \dotsc, \mathbf{x}_{k} \rbrace$, $U^{k} = \lbrace \mathbf{u}_{0}, \mathbf{u}_{1}, \dotsc, \mathbf{u}_{k} \rbrace$ and $Y^{k} = \lbrace \mathbf{y}_{0}, \mathbf{y}_{1}, \dotsc, \mathbf{y}_{k} \rbrace$ the state, control input and observations sequence, respectively. The control input $\mathbf{u}_{k-1}$ can be defined to influence the size of the measurement vector $\mathbf{y}_{k}$, its form, or both, and is selected by the controller based on the \textit{observation--control history} $\mathcal{F}_{k} = \sigma \lbrace Y^{k}, U^{k-1} \rbrace$, where $\sigma \lbrace z \rbrace$ represents the $\sigma$--algebra generated by $z$. We denote the finite set of all control inputs by $\mathcal{U} = \lbrace \mathbf{u}^{1}, \mathbf{u}^{2}, \dotsc, \mathbf{u}^{\alpha}\rbrace$.

\subsection{Review of Kalman--like Estimator}

In \cite{ZoisTSP14}, we developed an approximate nonlinear MMSE estimator for the Markov chain system state. This estimator is reviewed next. Let $\mathbf{p}_{k|k} \doteq [p_{k|k}^{1}, \dotsc, p_{k|k}^{n}]^{T} \in \mathcal{P} = \lbrace \mathbf{p} \in [0,1]^{n} ~|~ \mathbf{1}_{n}^{T}\mathbf{p} = 1 \rbrace$ denote the probability mass function (pmf) of $\mathbf{x}_{k}$ conditioned on $\mathcal{F}_{k}$ with $p_{k|k}^{i} = P(\mathbf{x}_{k} = \mathbf{e}_{i} | \mathcal{F}_{k}), \forall \mathbf{e}_{i} \in \mathcal{X}$. We have shown that this pmf (also known as \textit{belief state} \cite{BertsekasDPOC05}) coincides with the MMSE estimate of $\mathbf{x}_{k}$ given $\mathcal{F}_{k}$ and derived the following approximate MMSE estimator \cite{ZoisTSP14}.

\begin{thm}[\cite{ZoisTSP14}] \label{thm:approx_MMSE_estimate}
The Markov chain system estimate at time slot $k$ is recursively defined as
\begin{equation}\label{eq:filter_eqn}
\mathbf{p}_{k|k} = \mathbf{p}_{k|k-1} + \mathbf{G}_{k} [\mathbf{y}_{k} - \mathbf{y}_{k|k-1} ],~ k \geqslant 0
\end{equation}

\noindent
with
\begin{align}
\mathbf{p}_{k|k-1}& = \mathbf{P} \mathbf{p}_{k-1|k-1},\\
\mathbf{y}_{k|k-1}& = \mathcal{M}(\mathbf{u}_{k-1}) \mathbf{p}_{k|k-1},\\
\mathbf{G}_{k}& = \mathbf{\Sigma}_{k|k-1}\mathcal{M}^{T}(\mathbf{u}_{k-1})(\mathcal{M}(\mathbf{u}_{k-1}) \mathbf{\Sigma}_{k|k-1}\mathcal{M}^{T}(\mathbf{u}_{k-1}) + \widetilde{\mathbf{Q}}_{k})^{-1}
\end{align}

\noindent
where $\mathbf{p}_{0|-1} = \pi$, and $\pi$ is the initial distribution over the system states, $\mathcal{M}(\mathbf{u}_{k-1}) = [\mathbf{m}_{1}^{\mathbf{u}_{k-1}},\dotsc,\mathbf{m}_{n}^{\mathbf{u}_{k-1}}]$, $\mathbf{\Sigma}_{k|k-1}$ is the conditional covariance matrix of the prediction error and $\widetilde{\mathbf{Q}}_{k} = \sum_{i=1}^{n} p_{k|k-1}^{i} \mathbf{Q}_{i}^{\mathbf{u}_{k-1}}$.
\end{thm}

The proposed estimator is \textit{formally} similar to the Kalman filter but is a \underline{non--linear} estimator. Its MSE is given by the \textit{conditional filtering error covariance matrix} defined as
\begin{equation}\label{eq:cfecm}
\mathbf{\Sigma}_{k|k} \doteq \mathbb{E} \lbrace (\mathbf{x}_{k} - \mathbf{p}_{k|k}) (\mathbf{x}_{k} - \mathbf{p}_{k|k})^{T} | \mathcal{F}_{k} \rbrace = \diag(\mathbf{p}_{k|k}) - \mathbf{p}_{k|k}\mathbf{p}_{k|k}^{T}.
\end{equation}

\noindent
Since $\mathbf{p}_{k|k}$ is driven by control input selection, selecting the control sequence that minimizes the filter's MSE would result in good belief state estimates.

\subsection{Optimization Problem}\label{sec:OP}

\begin{figure}[h!]
\centering
\subfloat[Control input $\mathbf{u}^{1}$]
{\includegraphics[width=0.5\linewidth]{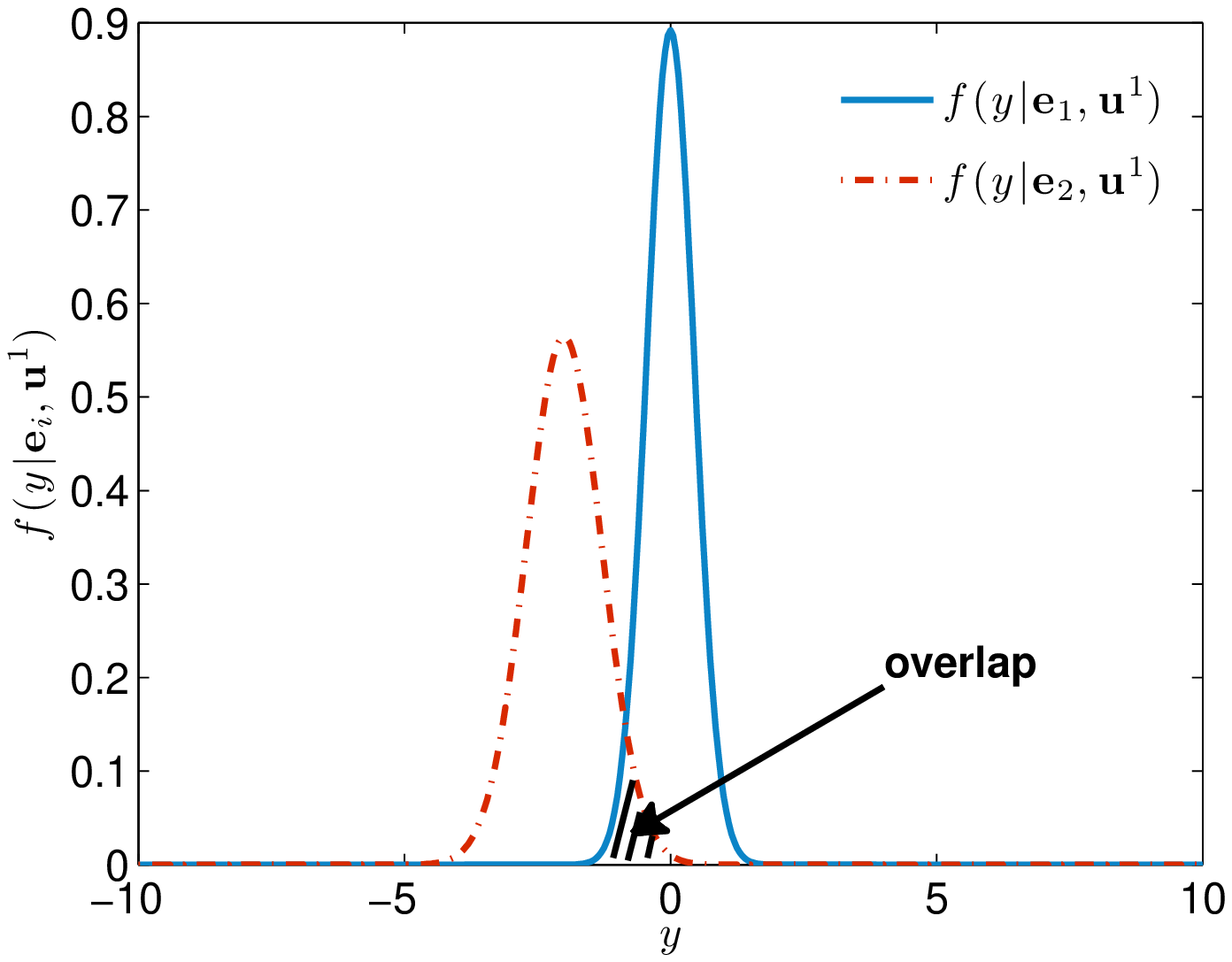}}
\subfloat[Control input $\mathbf{u}^{2}$]
{\includegraphics[width=0.5\linewidth]{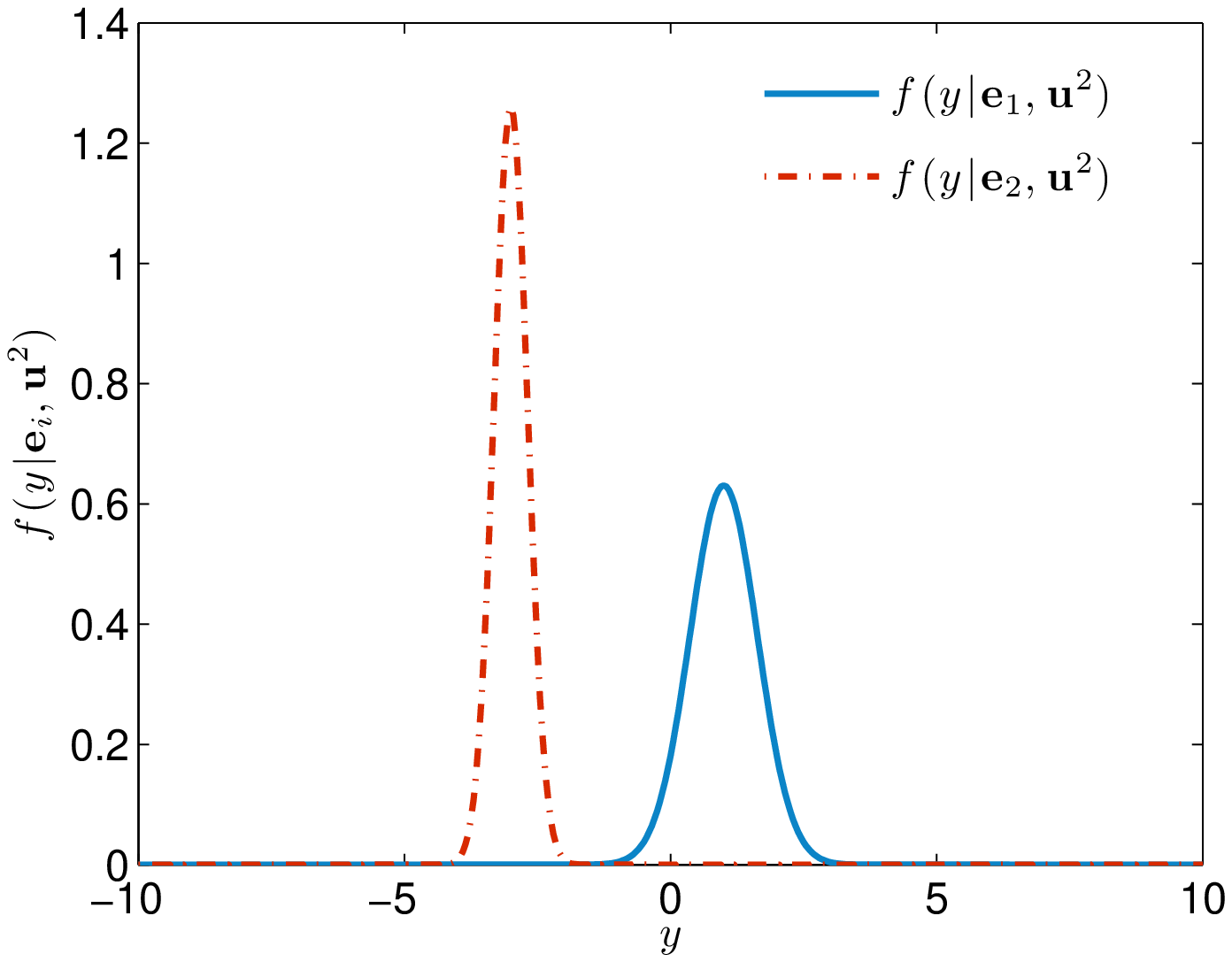}}
\caption{ Example of how control inputs $\mathbf{u}^{1}$ (\textit{left}) and $\mathbf{u}^{2}$ (\textit{right}) affect the observation kernel for states $\mathbf{e}_{1}$ and $\mathbf{e}_{2}$ resulting in errors due to overlap or not.}
\label{fig:control_effect}
\end{figure}

As shown in Fig.~\ref{fig:control_effect}, the proper choice of control input plays a crucial role in unveiling the true system state. For $n > 2$ states, selecting the appropriate control is complicated, since a control input that separates two states can bring closer any other two states. Furthermore, control input selection entails a usage cost, \emph{e.g.} power consumption spent for communicating certain number of samples from sensors to the fusion center. We are interested in two metrics: the \textit{estimation accuracy} and the \textit{sensing cost} associated with a certain control input. We underscore that different observations can provide better or worse qualitative views of the same system state, while incurring higher or lower sensing cost. We capture estimation accuracy by $\tr(\mathbf{\Sigma}_{k|k}(\mathbf{y}_{k},\mathbf{u}_{k-1})) \in [0,1]$, where the dependence of $\mathbf{\Sigma}_{k}$ on $\mathbf{y}_{k}$ and $\mathbf{u}_{k-1}$ has been stated explicitly. For each control input $\mathbf{u}_{k-1}$, the sensing cost is denoted by $c(\mathbf{u}_{k-1}) \in [0,1]$. To study the trade--off between estimation accuracy and energy consumption, we define the following objective function
\begin{equation}
g(\mathbf{y}_{k},\mathbf{u}_{k-1}) \doteq (1-\lambda) \tr(\mathbf{\Sigma}_{k|k}(\mathbf{y}_{k},\mathbf{u}_{k-1})) + \lambda c(\mathbf{u}_{k-1}),
\end{equation}

\noindent
where $\lambda \in [0,1]$. Next, we give a precise formulation of our active state tracking problem.

\vspace{5pt}
\noindent
\textbf{Active State Tracking Problem.} Under the stochastic system model given in Section \ref{subsec:SM}, our goal is to determine an admissible sensing strategy for the controller, \emph{i.e.} a sequence of control inputs $\mathbf{u}_{0}, \mathbf{u}_{1}, \dotsc, \mathbf{u}_{L-1}$, which solves for the following optimization problem
\begin{equation}\label{eq:optimization_problem}
\min_{\mathbf{u}_{0},\mathbf{u}_{1},\dotsc,\mathbf{u}_{L-1}} \mathbb{E} \bigg \lbrace \sum_{k=1}^{L} g(\mathbf{y}_{k},\mathbf{u}_{k-1}) \bigg \rbrace,
\end{equation}

\noindent
where  $L < \infty$ is the horizon length.

\section{Optimal Sensing Strategy}\label{sec:OSS}

The active state tracking problem introduced in Section \ref{sec:OP} constitutes a POMDP. The information $\mathcal{F}_{k}$ for decision making at time slot $k$ is of expanding dimension \cite{BertsekasDPOC05}. In contrast to standard POMDPs \cite{BertsekasDPOC05}, in our case, a memory--bounded sufficient statistic for decision making is the conditional distribution $\mathbf{p}_{k+1|k}$, which we refer to as \textit{predicted belief state} \cite{ZoisTSP14}. In one time step, its evolution follows Bayes' rule
\begin{equation}\label{eq:update_rule}
\mathbf{p}_{k+1|k} = \frac{\mathbf{P} \mathbf{r}(\mathbf{y}_{k}, \mathbf{u}_{k-1})\mathbf{p}_{k|k-1}}{\mathbf{1}_{n}^{T} \mathbf{r}(\mathbf{y}_{k}, \mathbf{u}_{k-1})\mathbf{p}_{k|k-1}} \doteq \mathbf{\Phi}(\mathbf{p}_{k|k-1},\mathbf{u}_{k-1},\mathbf{y}_{k}), 
\end{equation}

\noindent
where $\mathbf{r}(\mathbf{y}_{k}, \mathbf{u}_{k-1}) = \diag(f(\mathbf{y}_{k}|\mathbf{e}_{i},\mathbf{u}_{k-1}), \dotsc,f(\mathbf{y}_{k}|\mathbf{e}_{n},\allowbreak \mathbf{u}_{k-1}))$. The optimization problem formulated in (\ref{eq:optimization_problem}) can be solved using the finite--horizon DP equations given in Theorem \ref{thm:DP_algorithm} in terms of $\mathbf{p}_{k|k-1}$.

\begin{thm}\label{thm:DP_algorithm}
For $k = L-1, \dotsc, 1,$ the \textit{cost--to--go} function $\overline{J}_{k}(\mathbf{p}_{k|k-1})$ is related to $\overline{J}_{k+1}(\mathbf{p}_{k+1|k})$ through the recursion
\begin{align}\label{eq:DP_ss_bs_00}
\overline{J}_{k}(\mathbf{p}_{k|k-1}) = &\underset{\mathbf{u}_{k-1} \in \mathcal{U}}{\min} \bigg [ \ell(\mathbf{p}_{k|k-1},\mathbf{u}_{k-1}) + \int \mathbf{1}_{n}^{T} \mathbf{r}(\mathbf{y},\mathbf{u}_{k-1}) \mathbf{p}_{k|k-1} \overline{J}_{k+1}\bigg ( \frac{\mathbf{P} \mathbf{r}(\mathbf{y},\mathbf{u}_{k-1}) \mathbf{p}_{k|k-1}}{\mathbf{1}_{n}^{T}\mathbf{r}(\mathbf{y},\mathbf{u}_{k-1}) \mathbf{p}_{k|k-1}} \bigg ) d\mathbf{y} \bigg ],
\end{align}

\noindent
where $\ell(\mathbf{p}_{k|k-1},\mathbf{u}_{k-1}) = (1-\lambda)\mathbf{p}_{k|k-1}^{T} \mathbf{h}(\mathbf{p}_{k|k-1}, \mathbf{u}_{k-1}) + \lambda c(\mathbf{u}_{k-1})$ and $\mathbf{h}(\mathbf{p}_{k|k-1}, \mathbf{u}_{k-1})$ is a column vector with components $h(\mathbf{e}_{i},\mathbf{p}_{k|k-1},\mathbf{u}_{k-1}) = 1 - \tr{\big( \mathbf{G}_{k}^{T} \mathbf{G}_{k} \mathbf{Q}_{i}^{\mathbf{u}_{k-1}} \big)} -  \norm{\mathbf{p}_{k|k-1} + \mathbf{G}_{k} (\mathbf{m}_{i}^{\mathbf{u}_{k-1}} - \mathbf{y}_{k|k-1})}^2, i = 1, \dotsc, n$. The cost--to--go function for $k = L$ is given by
\begin{equation}\label{eq:DP_ss_bs_01}
\overline{J}_{L}(\mathbf{p}_{L|L-1}) = \underset{\mathbf{u}_{L-1} \in \mathcal{U}}{\min} \big [ \ell(\mathbf{p}_{L|L-1},\mathbf{u}_{L-1}) \big ].
\end{equation}
\end{thm}

\begin{IEEEproof}
For proof, see Appendix \ref{ap:DP_proof}.
\end{IEEEproof}

\noindent
\begin{rmk}
\emph{The cost functions in (\ref{eq:DP_ss_bs_00}) -- (\ref{eq:DP_ss_bs_01}) are \underline{non--linear} functions of the predicted belief state. Thus, the related POMDP is non--linear vis--\`a--vis standard POMDPs \cite{BertsekasDPOC05}.}
\end{rmk}

Solving the DP for a specific value of $\lambda$ yields the optimal sensing strategy for a given trade--off between estimation accuracy and sensing cost. However, the DP recursion does not directly translate to practical solutions due to the following issues: 1) the predicted belief state $\mathbf{p}_{k|k-1}$ is continuous valued, which implies that at each iteration, the cost--to--go function needs to be evaluated at each point of an uncountably infinite set, 2) the computation of the expected future cost requires a multi--dimensional integration, which is challenging, and 3) the non--linear form of the DP equations prevents the application of standard techniques \cite{SmallwoodOR73}, \cite{LovejoyAOR91}. We can still get an approximately optimal solution for small problem sizes by discretizing the space of predicted belief state estimates.

\section{Main Results}\label{sec:MR}

We next discuss structural properties of the cost--to-go function $\overline{J}_{k}(\cdot)$. We also exploit stochastic ordering \cite{MullerCMSMR02} to characterize the optimal sensing strategy in certain cases.

\subsection{Structural Properties}\label{subsec:SP}

We begin by simplifying the current cost $\ell(\mathbf{p}_{k|k-1},\mathbf{u}_{k-1})$, as shown in Lemma \ref{lem:current_cost_rw}.

\begin{lem}\label{lem:current_cost_rw}
The current cost $\ell(\mathbf{p}_{k|k-1},\mathbf{u}_{k-1})$ can be equivalently written as follows
\begin{align}
\ell(\mathbf{p}_{k|k-1},\mathbf{u}_{k-1})& = (1-\lambda)\tr{( (\mathbf{I} - \mathbf{G}_{k} \mathcal{M}(\mathbf{u}_{k-1} ) ) \mathbf{\Sigma}_{k|k-1} )} + \lambda c(\mathbf{u}_{k-1}).
\end{align}
\end{lem}
\begin{IEEEproof}
For proof, see Appendix \ref{ap:current_cost_rw}.
\end{IEEEproof}

Next, we state an important assumption that is necessary for proving the remaining results in this section.

\noindent
\begin{asmp}\label{asmp:2_states_1D}
\emph{We wish to distinguish between two system states, $\mathbf{e}_{1}$ and $\mathbf{e}_{2}$, using scalar measurements.}
\end{asmp}

Lemma \ref{lem:current_cost_rw} and Assumption \ref{asmp:2_states_1D} enable us to prove Lemma \ref{lem:concave_current_cost}, which we use to prove Theorem \ref{thm:concave_cost2go}.

\begin{lem}\label{lem:concave_current_cost}
Under Asssumption \ref{asmp:2_states_1D}, $\ell(\mathbf{p}_{k|k-1},\mathbf{u}_{k-1})$ is a concave function of the predicted belief state $\mathbf{p}_{k|k-1}$.
\end{lem}

\begin{IEEEproof}
For proof, see Appendix \ref{ap:concave_current_cost}.
\end{IEEEproof}

\begin{rmk}
\emph{Our numerical simulations imply that Lemma \ref{lem:concave_current_cost} holds for $n > 2$ states and multi--dimensional measurement vectors. However, due to the complicated expressions involved, we have yet to validate it analytically.}
\end{rmk}

\begin{thm}\label{thm:concave_cost2go}
Under Assumption 1, the cost--to--go function $\overline{J}_{k}(\mathbf{p}_{k|k-1}), k = L, L-1, \dotsc, 1,$ is a concave function of the predicted belief state $\mathbf{p}_{k|k-1}$.
\end{thm}

\begin{IEEEproof}
For proof, see Appendix \ref{ap:concave_cost2go}.
\end{IEEEproof}

\begin{figure}[h!]
\centering
  \includegraphics[width=0.7\linewidth]{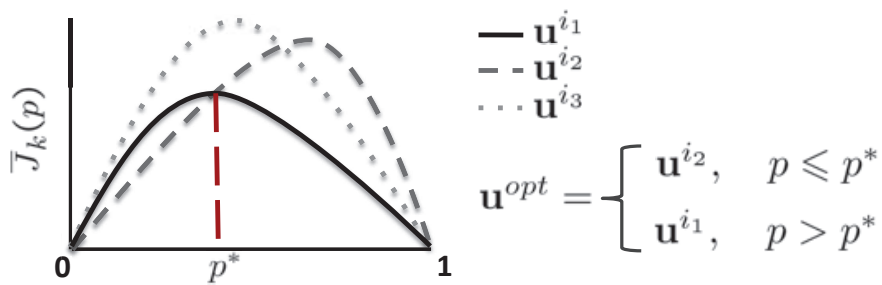}
  \caption{Optimal DP policy cost example for three control inputs and associated threshold sensing strategy rule.}\label{fig:ThresholdStructure}
\end{figure}

\noindent
A direct consequence of Theorem \ref{thm:concave_cost2go} is that the optimal sensing strategy has a threshold structure, which implies a very efficient implementation. Consider for example the scenario in Fig.~\ref{fig:ThresholdStructure}. Each line corresponds to the value of the term inside the minimization in (\ref{eq:DP_ss_bs_00}) for a different control input. Since the cost--to--go function is the minimum of these terms at each predicted belief state value, the intersection points correspond to decision thresholds that specify the change between control inputs. As a result, the optimal strategy reduces to testing in which interval the associated predicted belief state falls into and adopting the associated control input. This result for non--linear POMDPs generalizes the well--known fact that the optimal policy for linear POMDPs with two states has a threshold structure \cite{BertsekasDPOC05}. Note that, contrary to the non--linear POMDPs in \cite{KrishnamurthyTSP07}, we do not impose any constraints on the cost functions, Markov chain and observation probabilities to determine the optimality of the threshold structure. Finally, the concavity of the cost--to--go function enables us to characterize how informative a control input is, as we show in the sequel.

\subsection{Passive versus Active Sensing}\label{subsec:PvsAS}

A question of key interest is when a static or passive sensing policy is optimal. Herein, we exploit stochastic ordering of the observation kernels to characterize the structure of the optimal sensing strategy in several cases. According to Theorem \ref{thm:concave_cost2go}, for fixed control input $\mathbf{u}_{k-1}$, the cost--to--go function clearly depends on the observation kernel and the predicted belief state. Before, we proceed, we state the following definition.

\begin{defn} [Blackwell Ordering \cite{BlackwellAMS53}]
\emph{Given two conditional probability densities $f(\mathbf{y}|\mathbf{x}, \mathbf{u}^{a})$ and $f(\mathbf{y}|\mathbf{x}, \mathbf{u}^{b})$ from $\mathcal{X}$ to $\mathcal{Y}$, we say that $f(\mathbf{y}|\mathbf{x}, \mathbf{u}^{b})$ is \textit{less informative} than $f(\mathbf{y}|\mathbf{x}, \mathbf{u}^{a})$ $\big ( f(\mathbf{y}|\mathbf{x}, \mathbf{u}^{b}) \leqslant_{B} f(\mathbf{y}|\mathbf{x}, \mathbf{u}^{a}) \big )$ if there exists a \textit{stochastic transformation} $W$ from $\mathcal{Y}$ to $\mathcal{Y}$ such that $f(\mathbf{y}|\mathbf{x}, \mathbf{u}^{b}) = \int f(\mathbf{z}|\mathbf{x}, \mathbf{u}^{a}) \allowbreak W(\mathbf{z};\mathbf{y}) d \mathbf{z},\forall \mathbf{x} \in \mathcal{X}$.}
\end{defn}

\noindent
The following statement constitutes an important outcome of Blackwell ordering.

\begin{fct} [see \cite{DeGrootOSD70} ch. 14.17 and \cite{RiederMMOR91} Theorem 3.2]\label{fct:order_fcost}
Let $f(\mathbf{y}|\mathbf{x}, \mathbf{u}^{a})$ and $f(\mathbf{y}|\mathbf{x}, \mathbf{u}^{b})$ be two observation kernels. If $f(\mathbf{y}|\mathbf{x}, \mathbf{u}^{b}) \leqslant_{B} f(\mathbf{y}|\mathbf{x}, \mathbf{u}^{a})$, then $(\mathbb{T}_{a}g)(\mathbf{p}) \leqslant (\mathbb{T}_{b}g)(\mathbf{p}), \forall \mathbf{p} \in \mathcal{P}$ and for any concave function $g: \mathcal{P} \rightarrow \mathbb{R}$ with $(\mathbb{T}_{a}g)(\mathbf{p}) = \mathbb{E} \lbrace g(\mathbf{\Phi}(\mathbf{p},\mathbf{u}^{a},\mathbf{y})) \rbrace$, where expectation is with respect to $f(\mathbf{y}|\mathbf{x}, \mathbf{u}^{a})$.
\end{fct}

We restrict our attention to cases that satisfy Assumption \ref{asmp:2_states_1D} and to determine conditions that characterize the optimal control strategy structure, we consider the following four cases
\begin{enumerate}
\item[i.] \underline{Case I}: $m_{1}^{\mathbf{u}} = m_{2}^{\mathbf{u}}$ and $\sigma_{1,\mathbf{u}}^{2} = \sigma_{2,\mathbf{u}}^{2}$, $\mathbf{u} \in \mathcal{U}$,
\item[ii.] \underline{Case II}: $m_{1}^{\mathbf{u}} = m_{2}^{\mathbf{u}}$ and $\sigma_{1,\mathbf{u}}^{2} \neq \sigma_{2,\mathbf{u}}^{2}$, $\mathbf{u} \in \mathcal{U}$,
\item[iii.] \underline{Case III}: $m_{1}^{\mathbf{u}} \neq m_{2}^{\mathbf{u}}$ and $\sigma_{1,\mathbf{u}}^{2} = \sigma_{2,\mathbf{u}}^{2}$, $\mathbf{u} \in \mathcal{U}$,
\item[iv.] \underline{Case IV}: $m_{1}^{\mathbf{u}} \neq m_{2}^{\mathbf{u}}$ and $\sigma_{1,\mathbf{u}}^{2} \neq \sigma_{2,\mathbf{u}}^{2}$, $\mathbf{u} \in \mathcal{U}$.
\end{enumerate}

\noindent
Combining Fact \ref{fct:order_fcost} and Theorem \ref{thm:concave_cost2go} yields Corollary \ref{cor:passive_state_tracking}.

\begin{cor}\label{cor:passive_state_tracking}
Under Assumption \ref{asmp:2_states_1D} and for the active state tracking problem in (\ref{eq:optimization_problem}), if there exists a control input $\mathbf{u}^{*}$ satisfying $f(y|\mathbf{x},\mathbf{u}) \leqslant_{B} f(y|\mathbf{x},\mathbf{u}^{*})$ and $\ell(\mathbf{p},\mathbf{u}) \geqslant \ell(\mathbf{p},\mathbf{u}^{*}), \forall \mathbf{u} \in \mathcal{U}, \forall \mathbf{p} \in \mathcal{P}$, it is always optimal to select control input $\mathbf{u}^{*}$ irrespectively of the predicted belief state $\mathbf{p}$.
\end{cor}

\begin{figure}[h!]
\centering
\subfloat[][]{\includegraphics[width=0.5\linewidth]{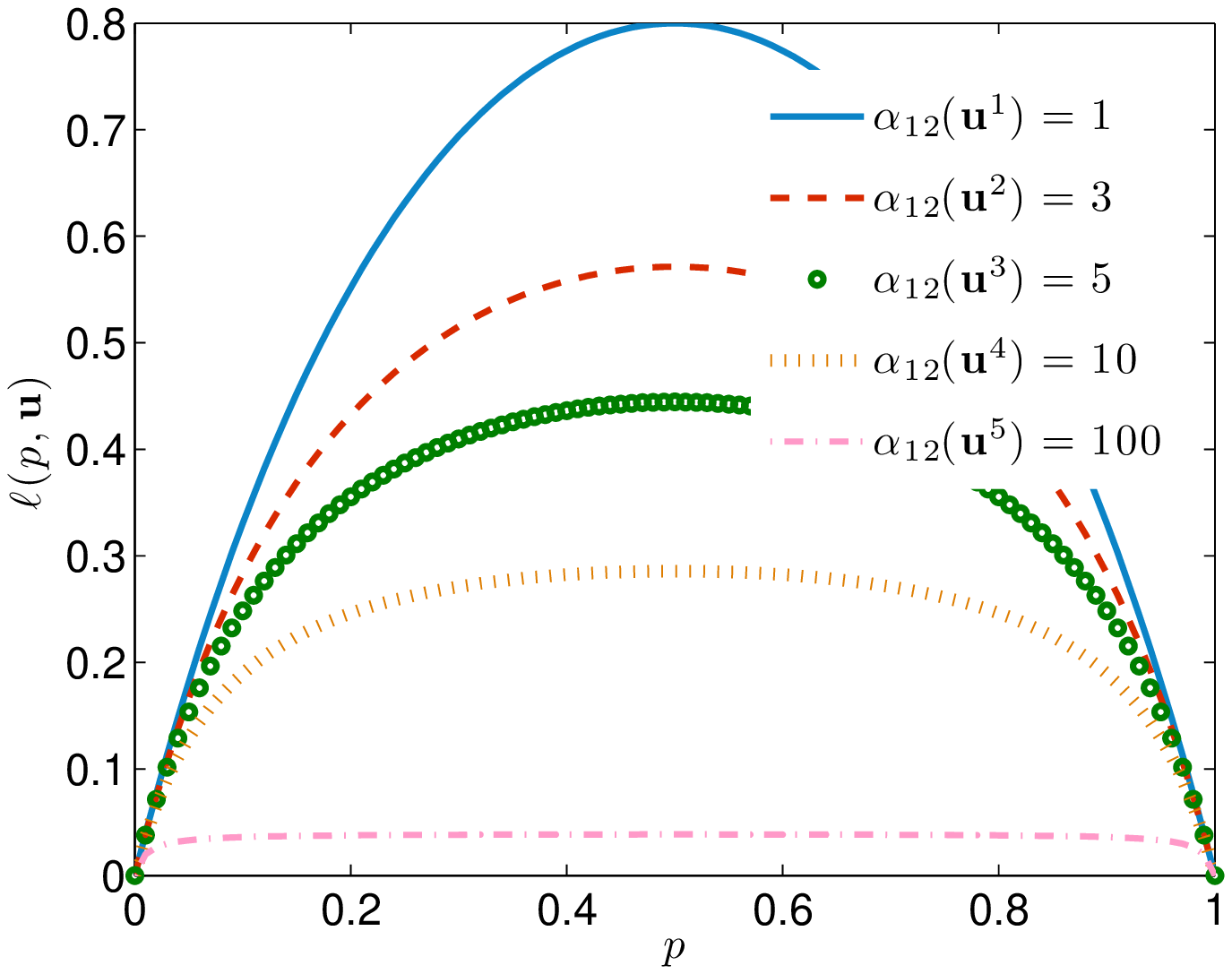}\label{fig:Case3}}
\subfloat[][]{\includegraphics[width=0.53\linewidth]{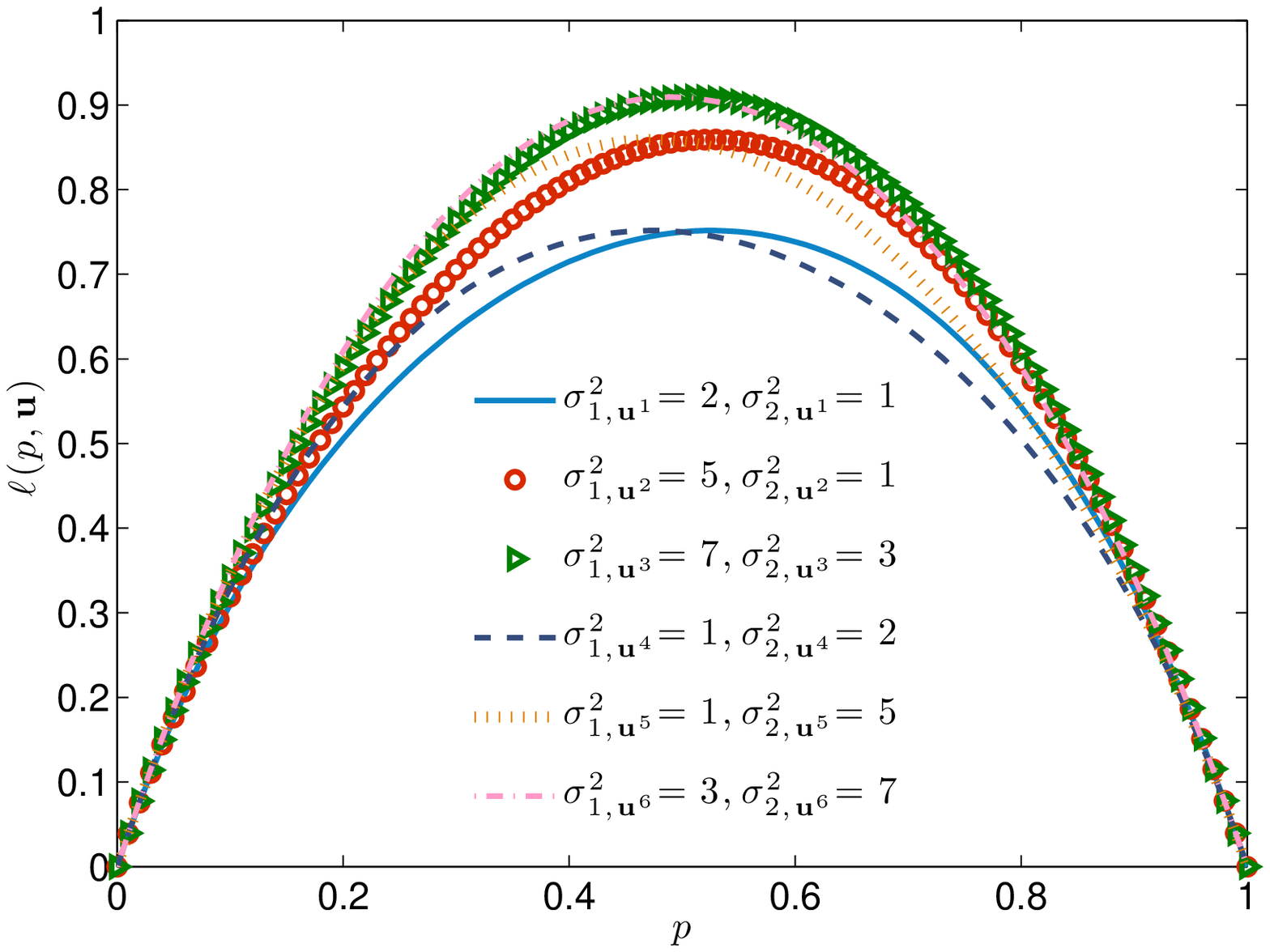}
\label{fig:Case4}}
\caption{\protect\subref{fig:Case3} Current costs for fixed variance $\sigma_{\mathbf{u}^{i}}^{2} = 2$ and different $a_{12}(\mathbf{u}^{i})$. \protect\subref{fig:Case4} Current costs with different variances and $a_{12}(\mathbf{u}^{i}) = constant$.}\label{fig:ActiveVSPassive}
\end{figure}

\noindent
Corollary \ref{cor:passive_state_tracking} provides a set of sufficient conditions for reducing active state tracking to passive state tracking with no observation control. For Cases I and II, we note that the current cost depends on the sensing cost associated with a certain control input, \emph{i.e.} $\ell(p,\mathbf{u}) = 2(1-\lambda)p(1-p) + \lambda c(\mathbf{u})$. If we were to order all controls with respect to the current cost only, then: $\ell(p,\mathbf{u}^{a}) \leqslant \ell(p,\mathbf{u}^{b}) \Leftrightarrow c(\mathbf{u}^{a}) \leqslant c(\mathbf{u}^{b}), \forall p \in \mathcal{P}$. Thus, we need to consider both the sensing costs of the controls and the Blackwell ordering of the related observation kernels to determine the optimal control input. Furthermore, under Assumption \ref{asmp:2_states_1D} and for Case II, the Blackwell ordering coincides with the ordering of the associated variances \cite{RiederMMOR91}, \emph{i.e.} $\sigma_{1,\mathbf{u}^{b}}^{2} \geqslant\sigma_{1,\mathbf{u}^{a}}^{2} \Rightarrow f(y|\mathbf{x},\mathbf{u}^{b}) \leqslant_{B} f(y|\mathbf{x},\mathbf{u}^{a})$. In Case III, the current cost has the form
\begin{equation}\label{eq:diff_means_same_vars}
\ell(p,\mathbf{u}) =  (1-\lambda)\frac{2\sigma_{\mathbf{u}}^{2}f(p)}{a_{12}(\mathbf{u})f(p) + \sigma_{\mathbf{u}}^{2}} + \lambda c(\mathbf{u}),
\end{equation}

\noindent
and for $\lambda = 0$, ordering the related costs can be achieved based on $a_{12}(\mathbf{u}) = (m_{1}^{\mathbf{u}}-m_{2}^{\mathbf{u}})^2$, as visually verified in Fig.~\ref{fig:Case3}. Corollary \ref{cor:diff_means_same_vars} gives more general conditions under which this ordering can be achieved.

\begin{cor}\label{cor:diff_means_same_vars}
Under Assumption \ref{asmp:2_states_1D} and for control inputs $\mathbf{u}^{i}$, $\mathbf{u}^{j} \in \mathcal{U}$, if either of the two conditions 
\begin{enumerate}
\item[C1)] $c(\mathbf{u}^{i}) = c(\mathbf{u}^{j})$,
\item[C2)] $a(\mathbf{u}^{i}) > a(\mathbf{u}^{j})$ and $c(\mathbf{u}^{i}) < c(\mathbf{u}^{j})$,
\end{enumerate}

\noindent
are met, $\mathbf{u}^{i}$ gives rise to the smallest current cost irrespective of the predicted belief state $p$.
\end{cor}

\begin{IEEEproof}
For proof, see Appendix \ref{ap:diff_means_same_vars}.
\end{IEEEproof}

For the more general Case IV, selecting the optimal control input is not straightforward. In fact, it depends on the predicted belief state, as Corollary \ref{cor:CaseIV} reveals and Fig.~\ref{fig:Case4} illustrates.

\begin{cor}\label{cor:CaseIV}
Under Assumption \ref{asmp:2_states_1D} and for two control inputs $\mathbf{u}^{a}$ and $\mathbf{u}^{b}$ with $a_{12}(\mathbf{u}^{a}) = a_{12}(\mathbf{u}^{b})$, $c(\mathbf{u}^{a}) = c(\mathbf{u}^{b})$, $\sigma_{1,\mathbf{u}^{a}}^{2} > \sigma_{1,\mathbf{u}^{b}}^{2}$ and $\sigma_{2,\mathbf{u}^{a}}^{2} < \sigma_{2,\mathbf{u}^{b}}^{2}$, there exists $p^{*} \in \mathcal{P}$ such that for $p \leqslant p^{*}, \ell(p,\mathbf{u}^{a}) \leqslant \ell(p,\mathbf{u}^{b})$ and for $p \geqslant p^{*}, \ell(p,\mathbf{u}^{a}) \geqslant \ell(p,\mathbf{u}^{b})$ with $p^{*} = \frac{\sigma_{2,\mathbf{u}^{b}}^{2} - \sigma_{2,\mathbf{u}^{a}}^{2}}{\sigma_{1,\mathbf{u}^{a}}^{2} - \sigma_{1,\mathbf{u}^{b}}^{2} + \sigma_{2,\mathbf{u}^{b}}^{2} - \sigma_{2,\mathbf{u}^{a}}^{2}}$.
\end{cor}

\begin{IEEEproof}
For proof, see Appendix \ref{ap:CaseIV}.
\end{IEEEproof}

\noindent
Intuitively, fixing $a_{12}(\mathbf{u}^{i})$ and increasing the associated variances leads to larger cost. Based on the above observations, for Case IV, active sensing is unavoidable, and the associated thresholds constitute a complicated function of the related means, variances and sensing costs.

\section{Low Complexity Strategies}\label{sec:LCS}

In this section, we propose two sensing strategies with lower complexity and discuss their implementation.

\subsection{Myopic Strategy}\label{subsec:MS}

Starting from the DP recursion in (\ref{eq:DP_ss_bs_00}), we propose a myopic algorithm that selects an appropriate control input by minimizing the one--step ahead cost, \emph{i.e.}
\begin{equation}\label{eq:myopic_strategy}
\mathbf{u}_{k}^{myopic} = \arg\min \ell(\mathbf{p}_{k+1|k},\mathbf{u}_{k}).
\end{equation}

\noindent
We note that the above solution avoids the computation of the expected future cost that requires a multi--dimensional integration. Still, the non--linear form of $\ell(\mathbf{p}_{k+1|k}, \mathbf{u}_{k})$ can be an issue. On the other hand, Lemma \ref{lem:concave_current_cost} implies an efficient implementation of the proposed algorithm in the case of two states and scalar measurements. We denote $q(\mathbf{p}_{k+1|k}) = \min_{\mathbf{u}_{k} \in \mathcal{U}} \ell(\mathbf{p}_{k+1|k},\mathbf{u}_{k})$. For each distinct $\mathbf{u}_{k}$, the function $\ell(\mathbf{p}_{k+1|k},\mathbf{u}_{k})$ is a concave function of $\mathbf{p}_{k+1|k}$ and this implies that $q(\mathbf{p}_{k+1|k})$ consists of segments of these concave functions. The last observation implies that for the setting in Lemma \ref{lem:concave_current_cost}, the myopic policy has a threshold structure of the form
\begin{equation}
\mathbf{u}_{k}^{myopic} = \left\{
  \begin{array}{l l}
    \mathbf{u}^{i_{1}} &, ~0 \leqslant p \leqslant p_{i_{1}}^{*},\\
    \mathbf{u}^{i_{2}} &, ~p_{i_{1}}^{*} < p \leqslant p_{i_{2}}^{*},\\
    ~\vdots &, \quad \quad \quad \vdots \\
    \mathbf{u}^{i_{J}} &, ~p_{i_{\Xi}}^{*} < p \leqslant p_{i_{\Xi+1}}^{*},\\
  \end{array} \right.
\end{equation}

\noindent
where $\Xi + 1$ denotes the number of different thresholds. Note that it is possible for a function $\ell(\mathbf{p}_{k+1|k},\mathbf{u}_{k})$ not to participate at all in $q(\mathbf{p}_{k+1|k})$ and in practice, a few number of them participate in $q(\mathbf{p}_{k+1|k})$. The threshold structure of the policy enables the following implementation: \textit{examine in which interval the predicted belief state falls into and declare as sensing choice, the associated control input.} As already discussed, this holds also true for the optimal sensing strategy.

\subsection{CE--WWLB Strategy}\label{subsec:E2WWLB}

As already discussed in Section \ref{sec:PD}, we are interested in optimizing the trade--off between estimation accuracy and sensing usage cost. In this section, we propose a sensing strategy that exploits a lower bound on the MSE in an effort to acquire a computationally efficient algorithm.

\subsubsection{Weiss--Weinstein Lower Bound}

The WWLB \cite{WeinsteinIT88, VanTrees07} is a Bayesian bound on the MSE, where the parameters of interest are random variables with known \emph{\`a priori} distribution. Consider $\bm{\theta} \in \mathbb{R}^{\ell}$ to be a random vector of parameters and $\mathbf{z} \in \mathbb{R}^{m}$ an associated measurement vector. Then, for any estimator $\hat{\bm{\theta}}(\mathbf{z})$, the error covariance matrix satisfies the inequality
\begin{equation}\label{eq:WWB}
\mathbb{E} \lbrace (\bm{\theta} - \hat{\bm{\theta}}(\mathbf{z}))(\bm{\theta} - \hat{\bm{\theta}}(\mathbf{z}))^{T} \rbrace \geqslant \mathbf{H} \mathbf{G}^{-1} \mathbf{H}^{T},
\end{equation}

\noindent
where $\mathbf{H} = [\mathbf{h}_{1},\mathbf{h}_{2},\dotsc,\mathbf{h}_{\ell}] \in \mathbb{R}^{\ell \times \ell}$ is a matrix with columns $\mathbf{h}_{i}, i = 1, \dotsc, \ell$, representing different ``test point" vectors, the $(i,j)$ element of matrix $\mathbf{G}$ is given by 
\begin{equation}\label{eq:Gmatrix}
[\mathbf{G}]_{ij} = \frac{\mathbb{E} \bigg \lbrace \bigg ( L^{s_{i}} (\mathbf{z}; \bm{\theta} + \mathbf{h}_{i},\bm{\theta}) - L^{1 - s_{i}} (\mathbf{z}; \bm{\theta} - \mathbf{h}_{i},\bm{\theta}) \bigg ) \bigg ( L^{s_{j}} (\mathbf{z}; \bm{\theta} + \mathbf{h}_{j},\bm{\theta}) - L^{1 - s_{j}} (\mathbf{z}; \bm{\theta} - \mathbf{h}_{j},\bm{\theta}) \bigg ) \bigg \rbrace }{ \mathbb{E} \bigg \lbrace L^{s_{i}} (\mathbf{z}; \bm{\theta} + \mathbf{h}_{i},\bm{\theta}) \bigg \rbrace \mathbb{E} \bigg \lbrace L^{s_{j}} (\mathbf{z}; \bm{\theta} + \mathbf{h}_{j},\bm{\theta}) \bigg \rbrace}
\end{equation}

\noindent
for any set of numbers $s_{i} \in (0,1)$ and $L(\mathbf{z};\bm{\theta}_{1},\bm{\theta}_{2}) = \frac{p(\mathbf{z},\bm{\theta}_{1})}{p(\mathbf{z},\bm{\theta}_{2})}$ is the joint likelihood ratio. Eq.~(\ref{eq:Gmatrix}) indicates that the matrix $\mathbf{G}$ is symmetric. Also, the matrix $\mathbf{H}$ and the set of numbers $\lbrace s_{1}, s_{2}, \dotsc, s_{\ell} \rbrace$ are arbitrary, \emph{i.e.} (\ref{eq:WWB}) represents a family of estimation error bounds. The choice $s_{i} = \frac{1}{2}, i = 1, 2, \dotsc, \ell,$ usually maximizes the WWLB \cite{WeinsteinIT88}. Furthermore, the test points avoid the regularity conditions imposed by other well--known bounds \cite{VanTrees07}. As a result, the WWLB can be applied to various cases, where the traditional bounds cannot, \emph{i.e.} in the estimation of discrete parameters for our problem of interest.

The sequential WWLB is an extension of the WWLB for Markovian dynamical systems \cite{ReeceFusion05, RapoportTSP07, XaverTSP13}. Specifically, let $\mathbf{H}_{k}$ and $\mathbf{G}_{k}$ be the matrices defined above calculated for $X^{k}$, $Y^{k}$ and $U^{k}$. To enable a sequential calculation of the WWLB, the matrix $\mathbf{H}_{k} = \blkdiag(\mathbf{H}_{0,0},\mathbf{H}_{1,1},\dotsc,\mathbf{H}_{k,k})$, where the submatrix $\mathbf{H}_{r,r} = [\mathbf{h}_{r}^{1},\mathbf{h}_{r}^{2},\dotsc,\mathbf{h}_{r}^{\ell}]$ refers to the state vector $\mathbf{x}_{r}$. We set $s_{i} = \frac{1}{2}, i = 1, 2, \dotsc, \ell$. Then, the sequential WWLB at time step $k$ is \cite{ReeceFusion05, RapoportTSP07, XaverTSP13}
\begin{equation}\label{eq:SWWB}
\mathbb{E} \big \lbrace (\mathbf{x}_{k} - \hat{\mathbf{x}}_{k|k})(\mathbf{x}_{k} - \hat{\mathbf{x}}_{k|k})^{T} \big \rbrace \geqslant \mathbf{H}_{k,k} \mathbf{J}_{k}^{-1} \mathbf{H}_{k,k}^{T},
\end{equation}

\noindent
where $\hat{\mathbf{x}}_{k|k}$ is an estimator of system state $\mathbf{x}_{k}$. The information submatrix $\mathbf{J}_{k+1}$ is recursively updated as follows \cite{ReeceFusion05, RapoportTSP07, XaverTSP13}
\begin{align}
\mathbf{A}_{k+1}& = \mathbf{G}^{k+1}_{k,k} - \mathbf{G}^{k}_{k,k-1} \mathbf{A}_{k}^{-1} \mathbf{G}^{k}_{k-1,k}, \label{eq:recursive_update_00} \\
\mathbf{J}_{k+1}& = \mathbf{G}^{k+1}_{k+1,k+1} - \mathbf{G}^{k+1}_{k+1,k} \mathbf{A}_{k+1}^{-1} \mathbf{G}^{k+1}_{k,k+1},\label{eq:recursive_update_01}
\end{align}

\noindent
$\forall k = 0, 1, \dotsc, $ where $\mathbf{G}^{k+1}_{i,j} \in \mathbb{R}^{\ell \times \ell}$ and $\mathbf{G}^{k}_{e,f} \in \mathbb{R}^{\ell \times \ell}$ are entries of the matrices $\mathbf{G}_{k+1}$ and $\mathbf{G}_{k}$, respectively. Due to the symmetry of $\mathbf{G}_{k+1}$ and $\mathbf{G}_{k}$, we have that 1) $\mathbf{G}^{k+1}_{i,j} = \mathbf{G}^{k+1}_{j,i}$, and 2) $\mathbf{G}^{k}_{e,f} = \mathbf{G}^{k}_{e,f}$. Matrices $\mathbf{A}_{0}^{-1} \doteq \mathbf{0}$, $\mathbf{G}_{0,-1}^{0} \doteq \mathbf{0}$, $\mathbf{G}_{-1,0}^{0} \doteq \mathbf{0}$ and $\mathbf{J}_{0}^{-1}$ is the covariance matrix associated with $P(\mathbf{x}_{0}) f(\mathbf{y}_{0}|\mathbf{x}_{0}, \mathbf{u}_{-1})$, where $\mathbf{u}_{-1}$ is a fixed control input. Lemma \ref{lem:SWWLB_formulae} provides the exact form of the sequential WWLB for our system model.

\begin{lem}\label{lem:SWWLB_formulae}
For the system model described in Section \ref{subsec:SM}, let $P(x_{0})$ be the known \`a priori pmf related to the initial state $x_{0}$. Then, the sequential WWLB at each time step $k$ is determined by (\ref{eq:recursive_update_00}) and (\ref{eq:recursive_update_01}), where
\begin{align}
&G_{k+1,k+1}^{k+1} = \frac{2 \big ( 1 - \exp{(\eta_{k}(h_{k+1},-h_{k+1}))} \big )}{\exp{(2\eta_{k}(h_{k+1},0))}},\label{eq:Gmatrix_recursive_00}\\
&G_{k+1,k}^{k+1} = G_{k,k+1}^{k+1} = \frac{\exp(\zeta_{k}(h_{k},h_{k+1})) - \exp(\zeta_{k}(-h_{k},h_{k+1}))}{\exp(\eta_{k}(h_{k+1},0) + \rho_{k}(h_{k},0))} +\frac{\exp(\zeta_{k}(-h_{k},-h_{k+1}))-\exp(\zeta_{k}(h_{k},-h_{k+1}))}{\exp(\eta_{k}(,h_{k+1},0) + \rho_{k}(h_{k},0))},\label{eq:Gmatrix_recursive_01} \\
&G_{k,k}^{k+1} = \frac{2 \big ( 1 - \exp{(\rho_{k}(h_{k},-h_{k}))} \big )}{\exp{(2\rho_{k}(h_{k},0))}},\label{eq:Gmatrix_recursive_02}
\end{align}

\noindent
with
\begin{align}
\eta_{k}(h_{a},h_{b})& = \ln \sum_{x_{k}} P(x_{k}) \sum_{x_{k+1}} \sqrt{P(x_{k+1} + h_{a} | x_{k})} \sqrt{P(x_{k+1} + h_{b} | x_{k})}\xi(x_{k+1} + h_{a}, x_{k+1} + h_{b}), \label{eq:eta_term} \\
\rho_{k}(h_{a},h_{b})& = \ln \sum_{x_{k-1}} P(x_{k-1}) \sum_{x_{k}} \sqrt{P(x_{k} + h_{a}|x_{k-1})}\sqrt{P(x_{k} + h_{b}|x_{k-1})} \sum_{x_{k+1}}
\sqrt{P(x_{k+1} | x_{k} + h_{a})} \nonumber \\ &
\times \sqrt{P(x_{k+1} | x_{k} + h_{b})} \xi(x_{k} + h_{a}, x_{k} + h_{b}), \label{eq:rho_term} \\
\zeta_{k}(h_{a},h_{b})& = \ln \sum_{x_{k-1}} P(x_{k-1}) \sum_{x_{k}} \sqrt{P(x_{k} + h_{a} | x_{k-1}) P(x_{k} | x_{k-1})}\sum_{x_{k+1}} \sqrt{P(x_{k+1} | x_{k} + h_{a}) P(x_{k+1} + h_{b} | x_{k})} \nonumber \\ & \times \xi(x_{k} + h_{a}, x_{k}) \xi(x_{k+1} + h_{b}, x_{k+1}),\label{eq:zeta_term}
\end{align}

\noindent
and the function $\xi(\cdot,\cdot)$ corresponds to the Bhattacharyya coefficient given by \cite{Duda01}
\begin{align}\label{eq:Bh_coeff_final}
\xi(x_{k} + h_{a}, x_{k} + h_{b}) = \exp \bigg (  &- \bigg [ \frac{1}{8} \big( \mathbf{m}_{x_{k}+h_{a}}^{\mathbf{u}_{k-1}} - \mathbf{m}_{x_{k}+h_{b}}^{\mathbf{u}_{k-1}} \big )^{T} \mathbf{Q}_{h}^{-1}\big( \mathbf{m}_{x_{k}+h_{a}}^{\mathbf{u}_{k-1}} - \mathbf{m}_{x_{k}+h_{b}}^{\mathbf{u}_{k-1}} \big ) \nonumber \\
&+ \frac{1}{2} \log \frac{\det \mathbf{Q}_{h} }{\sqrt{\det \mathbf{Q}_{x_{k}+h_{a}}^{\mathbf{u}_{k-1}} \cdot \det \mathbf{Q}_{x_{k}+h_{b}}^{\mathbf{u}_{k-1}}}} \bigg ]\bigg ),
\end{align}

\noindent
where $2\mathbf{Q}_{h} = \mathbf{Q}_{x_{k}+h_{a}}^{\mathbf{u}_{k-1}} + \mathbf{Q}_{x_{k}+h_{b}}^{\mathbf{u}_{k-1}}$. Furthermore, the information submatrix $J_{0} = \frac{2 \big ( 1 - \exp{(\gamma(h_{0},-h_{0}))} \big )}{\exp{(2\gamma(h_{0},0))}}$ with $\gamma(h_{a},h_{b}) = \ln \sum_{{x}_{0}} \sqrt{P(x_{0} + h_{a}) P(x_{0} + h_{b})} \xi(x_{0} + h_{a}, x_{0} + h_{b})$.
\end{lem}

\begin{IEEEproof}
For proof, see Appendix \ref{ap:SWWLB_formulae_proof}.
\end{IEEEproof}

\noindent
\begin{rmk}
For our discrete--time, finite--state Markov chain with $n$ states\footnote{We have adopted the scalar notation $x_{k} \in \mathcal{X} \doteq \lbrace 1, \dotsc, n \rbrace$ to represent the system state at time step $k$.}, all variables in (\ref{eq:recursive_update_00}) and (\ref{eq:recursive_update_01}) are scalars.
\end{rmk}

As already discussed, the WWLB avoids the need to satisfy any regularity conditions via the usage of test points. For our system model, this fact implies that we can determine the exact form of the sequential WWLB through Lemma \ref{lem:SWWLB_formulae}. Nonetheless, the test points must be carefully selected to account for the fact that our parameter space is discrete. In other words, test points should be state--dependent, \emph{i.e.} $h_{t} \in \mathcal{A} \doteq \big \lbrace h_{t}(x_{t}) \in \mathbb{R} ~|~ x_{t} + h_{t}(x_{t}) \in \mathcal{X} \big \rbrace$ to ensure the validity and correctness of all related formulae. For instance, for $n = 4$ states $\lbrace 1, 2, 3, 4 \rbrace$, the valid test point values for each state are: 1) $h_{t}(1) \in \lbrace 1, 2, 3 \rbrace,$ 2) $h_{t}(2) \in \lbrace -1, 1, 2 \rbrace,$ 3) $h_{t}(3) \in \lbrace -2, -1, 1 \rbrace,$ and 4) $h_{t}(4) \in \lbrace -3, -2, -1 \rbrace$.

\noindent
\begin{rmk}
The WWLB computed above assumes one test point per parameter, and can be easily extended to accommodate multiple test points per parameter \cite{VanTrees07}. This significantly increases the associated computational complexity, but in some cases, multiple test points are required to obtain a tight bound.
\end{rmk}

\subsubsection{Cost--Efficient WWLB (CE--WWLB)}

We propose the following strategy that optimizes the trade--off between the sequential WWLB and the sensing usage cost, \emph{i.e.}
\begin{equation}\label{eq:E2WWLB_strategy}
\mathbf{u}_{k}^{CE-WWLB} = \arg\min \big [ (1-\lambda) v(\mathbf{u}_{k}) + \lambda c(\mathbf{u}_{k}) \big ],
\end{equation}

\noindent
where $v(\mathbf{u}_{k}) \doteq \max_{\mathbf{h}_{k+1}} [ J_{k+1}^{-1}(\mathbf{h}_{k+1},\mathbf{u}_{k}) ]$, and the dependence of $J_{k+1}$ on $\mathbf{h}_{k+1}$ and $\mathbf{u}_{k}$ has been stated explicitly. The WWLB is maximized with respect to all possible test point combinations at each time step to ensure that the highest WWLB is computed.

Since the WWLB constitutes a lower bound on the MSE of any Markov chain system state estimator and we are interested in strategies that optimize the trade--off between MSE and sensing cost, the proposed strategy in (\ref{eq:E2WWLB_strategy}) is rather intuitive. Another agreeable characteristic is that the associated cost function $v(\mathbf{u}_{k})$ consists of functions of union--bound terms based on the Bhattacharyya detection error probability bound \cite{Duda01}. In fact, the terms in (\ref{eq:eta_term}) -- (\ref{eq:zeta_term}) can be expressed as functions of these bounds, \emph{e.g.} $\eta_{k}(h_{a},h_{b}) = \ln \sum_{x_{k}} P(x_{k}) P_{ub}^{Bh}(x_{k})$, where $P_{ub}^{Bh}(x_{k}) = \sum_{x_{k+1}} \sqrt{P(x_{k+1} + h_{a} | x_{k}) P(x_{k+1} + h_{b} | x_{k})} \allowbreak \xi(x_{k+1} + h_{a}, x_{k+1} + h_{b})$. This last step builds a nice connection between MSE and detection error performance. Note that several sensing strategies, which have been empirically shown to perform well, have focused on the optimization of the Bhattacharyya coefficient and the detection error probability union bounds \cite{ZoisTSP13}, since these are good measures of the confusability of different hypotheses. At this point, we underscore that the Bhattacharyya coefficient in (\ref{eq:Bh_coeff_final}) follows from setting $s = s_{i} = \frac{1}{2}, i = 1, 2, \dotsc, \ell$. If we wish to also optimize the WWLB with respect to $s$, the resulting WWLB formulae\footnote{The square root terms will also be replaced by powers of functions of $s$.} will instead depend on $\int f(\mathbf{y}_{k} | x_{k} + h_{a}, \mathbf{u}_{k-1})^{s} f(\mathbf{y}_{k} | x_{k} + h_{b}, \mathbf{u}_{k-1})^{1-s} d \mathbf{y}_{k} = \exp(-\kappa(s))$, where
\begin{align}\label{eq:Chernoff_bound_error_exponent}
\kappa(s) \doteq & \frac{1}{2} \ln \frac{|s \mathbf{Q}_{x_{k} + h_{a}}^{\mathbf{u}_{k-1}} + (1-s) \mathbf{Q}_{x_{k} + h_{b}}^{\mathbf{u}_{k-1}}|}{|\mathbf{Q}_{x_{k} + h_{a}}^{\mathbf{u}_{k-1}}|^{s} |\mathbf{Q}_{x_{k} + h_{b}}^{\mathbf{u}_{k-1}}|^{1-s}} + \frac{s(1-s)}{2} \bigg( \mathbf{m}_{x_{k} + h_{b}}^{\mathbf{u}_{k-1}} - \mathbf{m}_{x_{k} + h_{a}}^{\mathbf{u}_{k-1}} \bigg)^{T} \bigg( s \mathbf{Q}_{x_{k} + h_{a}}^{\mathbf{u}_{k-1}} + (1-s) \mathbf{Q}_{x_{k} + h_{b}}^{\mathbf{u}_{k-1}} \bigg)^{-1} \nonumber \\ &\times \bigg( \mathbf{m}_{x_{k} + h_{b}}^{\mathbf{u}_{k-1}} -  \mathbf{m}_{x_{k} + h_{a}}^{\mathbf{u}_{k-1}} \bigg),
\end{align}

\noindent
that is the error exponent of the Chernoff bound \cite{Duda01}. In that case, the WWLB union--bound terms will be based on the Chernoff detection error probability bound \cite{Duda01}. Since the latter bound is tighter than the Bhattacharyya bound, the associated sensing strategy might lead to better trade--off curves than CE--WWLB, yet, with the expense of increased computational complexity due to the optimization over $s$. To avoid such an issue, we adopted the computationally simpler but slightly less tight Bhattacharyya bound.

The myopic structure of the proposed strategy in (\ref{eq:E2WWLB_strategy}) also benefits computational complexity, since the computational burden of determining the expected future cost required by the DP algorithm in (\ref{eq:DP_ss_bs_00}) is avoided. Furthermore, there is no need to consider every point of an uncountably infinite set, since the associated optimization function does not depend on the predicted belief state $\mathbf{p}_{k+1|k}$. Lastly, the WWLB constitutes an off--line performance bound, \emph{i.e.} the related measurement information is averaged out. As a result, off--line computation of this strategy is feasible.

\section{Numerical Results}\label{sec:NR}

\begin{figure}[h!]
\centering
  \includegraphics[width=0.6\linewidth]{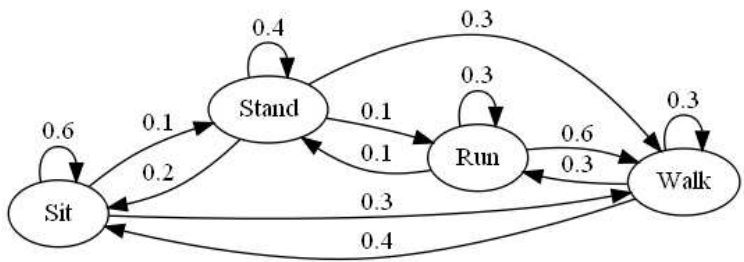}
  \caption{ A Markov chain of four physical activity states: \emph{Sit}, \emph{Stand}, \emph{Run}, \emph{Walk} \cite{ThatteTSP11}.}\label{fig:MarkovChainExample}
\end{figure}

\begin{figure}[h!]
\centering
  \includegraphics[width=0.8\linewidth]{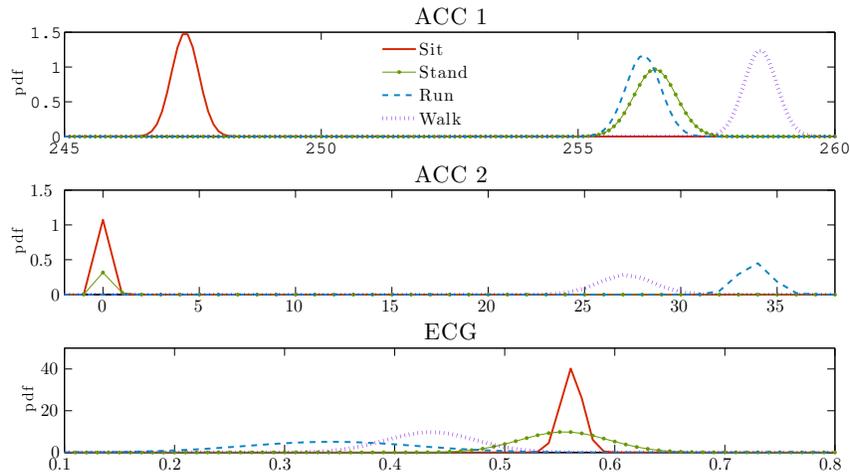}
  \caption{Signal model pdfs for four physical activity states and three biometric sensors for a single individual.}\label{fig:4Hypotheses}
\end{figure}

In this section, we illustrate the performance of the proposed sensing strategies in a body sensing application using real data \cite{ZoisTSP13}. We begin by introducing the body sensing problem. We consider an individual wearing a Wireless Body Area Network (WBAN), which consists of two accelerometers (ACCs), an electrocardiograph (ECG) and an energy--constrained mobile phone as a fusion center. The individual is changing between four physical activity states, \textit{Sit}, \textit{Stand}, \textit{Run} and \textit{Walk}, modeled by the discrete--time, finite--state Markov chain of Fig.~\ref{fig:MarkovChainExample}. At each time slot, a set of biometric signals is generated by the sensors, and feature extraction and selection techniques \cite{ThatteTSP11} are employed to produce a set of samples. In contrast to traditional sensor networks, where the sensors' energy--constrained nature impairs the network's lifetime, herein, continuously receiving samples from all the sensors limits the phone's battery life \cite{ZoisTSP13}. Meanwhile, the individual's physical activity state must be inferred at each time slot by appropriately using the information communicated by the biometric sensors. Thus, sensing strategies (such as the ones presented in Sections \ref{sec:OSS} and \ref{sec:LCS}) must be employed by the mobile phone to optimize the trade--off between estimation performance and energy consumption. Based on such strategies, the mobile phone can decide to receive all (or any subset) of the generated samples by selecting the appropriate control input $\mathbf{u}_{k} = [N_{1}^{\mathbf{u}_{k}}, N_{2}^{\mathbf{u}_{k}}, N_{3}^{\mathbf{u}_{k}}]^{T}$, where $N_{l}^{\mathbf{u}_{k}}$ denotes the total number of samples requested from sensor $S_{l}$ when control input $\mathbf{u}_{k}$ is selected. We assume that during each time slot $k$, there exists a fixed budget of $N$ samples that we cannot exceed, \emph{i.e.} $\mathbf{u}_{k}^{T} \mathbf{1} \leqslant N$, and the mobile phone can select between $\alpha = \sum_{i=1}^N \binom{i+2}{i}$ available measurement vectors of the form in (\ref{eq:observation_model}) with $\mathbf{m}_{i}^{\mathbf{u}_{k-1}}=[\bm\mu_{i,\mathbf{u}_{k-1}}(S_{1})^{T}, \bm\mu_{i,\mathbf{u}_{k-1}}(S_{2})^{T}, \bm\mu_{i,\mathbf{u}_{k-1}}(S_{3})^{T}]^T$ and $\mathbf{Q}_{i}^{\mathbf{u}_{k-1}} = \diag (\mathbfcal{Q}_{i,\mathbf{u}_{k-1}}(S_{1}), \mathbfcal{Q}_{i,\mathbf{u}_{k-1}}(S_{2}), \mathbfcal{Q}_{i,\mathbf{u}_{k-1}}(S_{3})),$ where $\bm\mu_{i,\mathbf{u}_{k-1}}(S_{l})$ is a $N_{l}^{\mathbf{u}_{k-1}} \times 1$ vector, $\mathbfcal{Q}_{i,\mathbf{u}_{k-1}}(S_{l}) = \frac{\sigma_{S_{l},i}^{2}}{1-\phi^2} \mathbf{T} + \sigma_{z}^{2} \mathbf{I}$ is a $N_{l}^{\mathbf{u}_{k-1}} \times N_{l}^{\mathbf{u}_{k-1}}$ matrix, $\mathbf{T}$ is a Toeplitz matrix whose first row/column is $[1, \phi , \phi^2 , \dotsc , \phi^{N_{l}^{\mathbf{u}_{k-1}} -1}]^{T}$, $\phi$ is the parameter of our model and $\sigma_{z}^{2}$ accounts for sensing and communication noise. The signal model pdfs for the four activity states and the three biometric sensors for a single individual are shown in Fig.~\ref{fig:4Hypotheses}. Finally, the sensing usage cost captures the \textit{normalized energy cost} $c(\mathbf{u}_{k}) \doteq \frac{1}{C} \mathbf{u}_{k}^{T} \bm \delta$, where $\bm \delta = [\delta_{\text{ACC 1}},\delta_{\text{ACC 2}},\delta_{\text{ECG}}]^{T} = [0.585, 0.776, 1]^{T}$ \cite{ZoisTSP13} is a vector that describes the mobile phone's reception cost for each of the biometric sensors, and $C$ is a normalizing factor.

Next, we compare the optimal sensing strategy of Theorem \ref{thm:DP_algorithm} (\emph{DP MSE--based strategy}) with the \emph{myopic strategy} of (\ref{eq:myopic_strategy}) and the \emph{CE--WWLB strategy} of (\ref{eq:E2WWLB_strategy}) with respect to: 1) the \emph{average MSE performance} defined as $\text{AMSE} \doteq \frac{1}{K} \sum_{k=1}^{K} \tr(\mathbf{\Sigma}_{k|k})$, 2) the \emph{average detection performance} defined as $\text{ADP} \doteq \frac{1}{K} \sum_{k=1}^{K} \mathbf{1}_{\lbrace \mathbf{x}_{k}=\hat{\mathbf{x}}_{k} \rbrace}$, where $\hat{\mathbf{x}}_{k} = \arg \max \mathbf{p}_{k|k}$, and 3) the \emph{average energy cost} defined as $\text{AEC} \doteq \frac{1}{K} \sum_{k=1}^{K} \mathbf{u}_{k}^{T} \bm \delta$, where $K$ represents the number of Monte Carlo runs. Unless stated, the simulation parameters are as follows: $N = 12$ samples in total, $L = 5$ and $K = 10^{6}$. We also compare with an \emph{equal allocation (EA)} strategy ($N = 3, 6, 9, 12$), where same number of samples are requested from each sensor irrespective of the individual's physical activity state.

\begin{figure}[h!]
\centering
\noindent
  \includegraphics[width=0.6\linewidth]{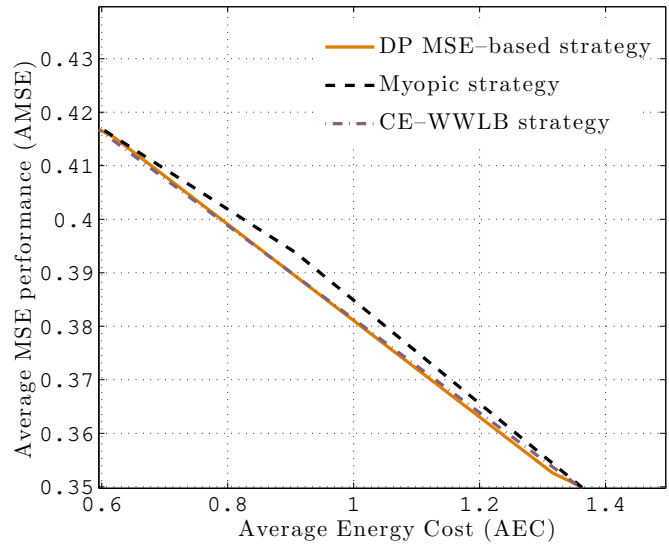}
  \caption{Trade--off curves for DP MSE--based, myopic and CE--WWLB strategies for $N = 2$ samples.}\label{fig:DP_vs_Myopic_vs_CEWWLB}
\end{figure}

Fig.~\ref{fig:DP_vs_Myopic_vs_CEWWLB} shows the AEC--AMSE trade--off curves of the DP MSE--based, myopic and CE--WWLB strategies. The total budget $N$ was set to two samples, since for $N > 2$, the optimal POMDP solution requires excessive amount of computation time. For these small problem sizes, the myopic and CE--WWLB strategies performed competitively with DP. In fact, the loss of performance due to adoption of myopic policy is small, while CE--WWLB's performance is essentially indistinguishable from the performance of the DP MSE--based strategy. Our intuition suggests that the WWLB successfully captures the detection nature of our active state tracking problem, which in turn justifies the suitability of functions of detection error probability bounds as performance objectives for this type of problems. Another agreeable characteristic of employing these strategies is the attendant complexity reduction, which is significant. Based on these findings, we increase the total number of samples $N$ to compare the lower complexity strategies, and remove the computationally intractable optimal method from further consideration.

\begin{figure}[h!]
\centering
\subfloat[AMSE versus AEC]{\includegraphics[width=0.53\linewidth]{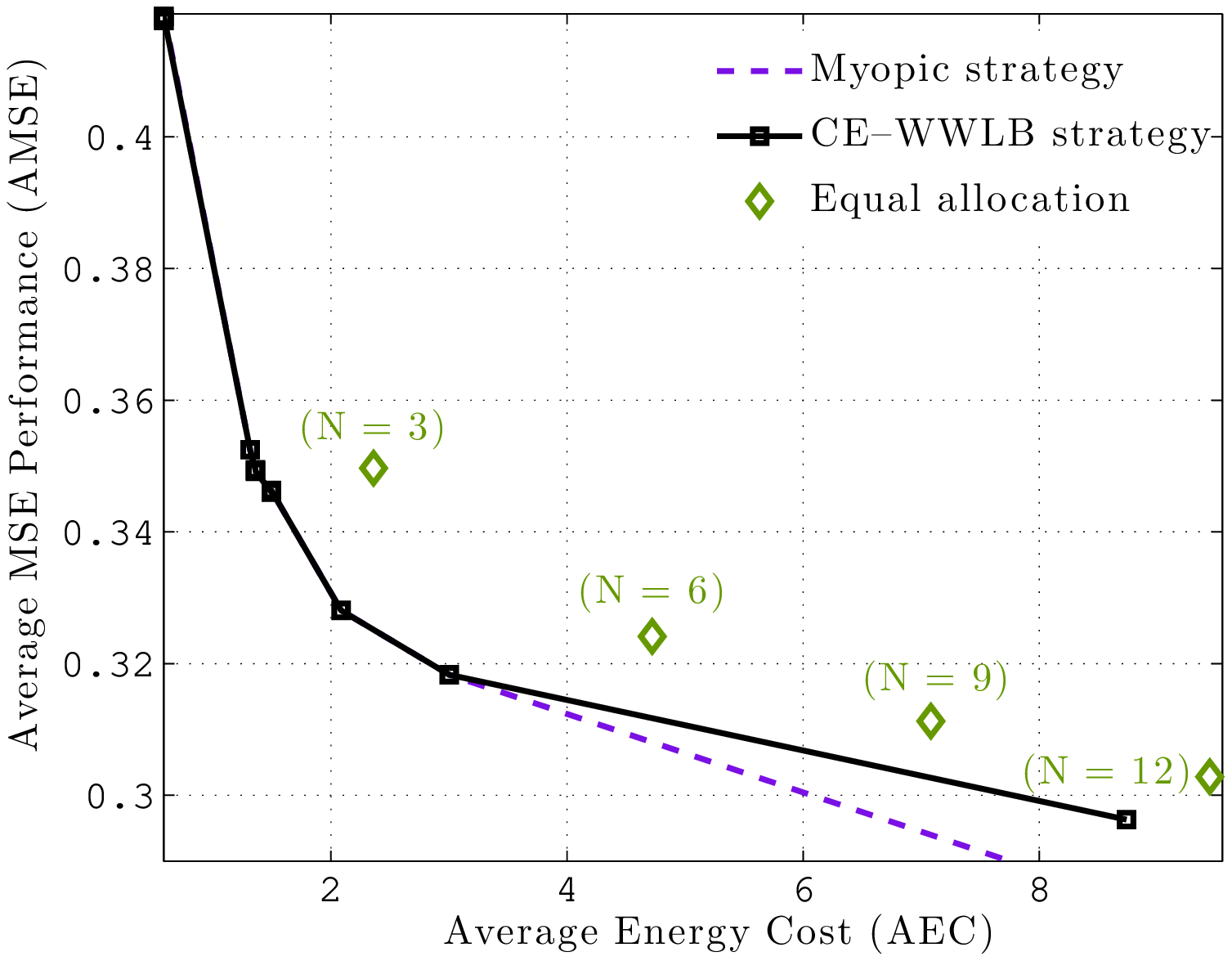}\label{fig:trade_off_curve_MSE}}
\subfloat[ADP versus AEC]{\includegraphics[width=0.5\linewidth]{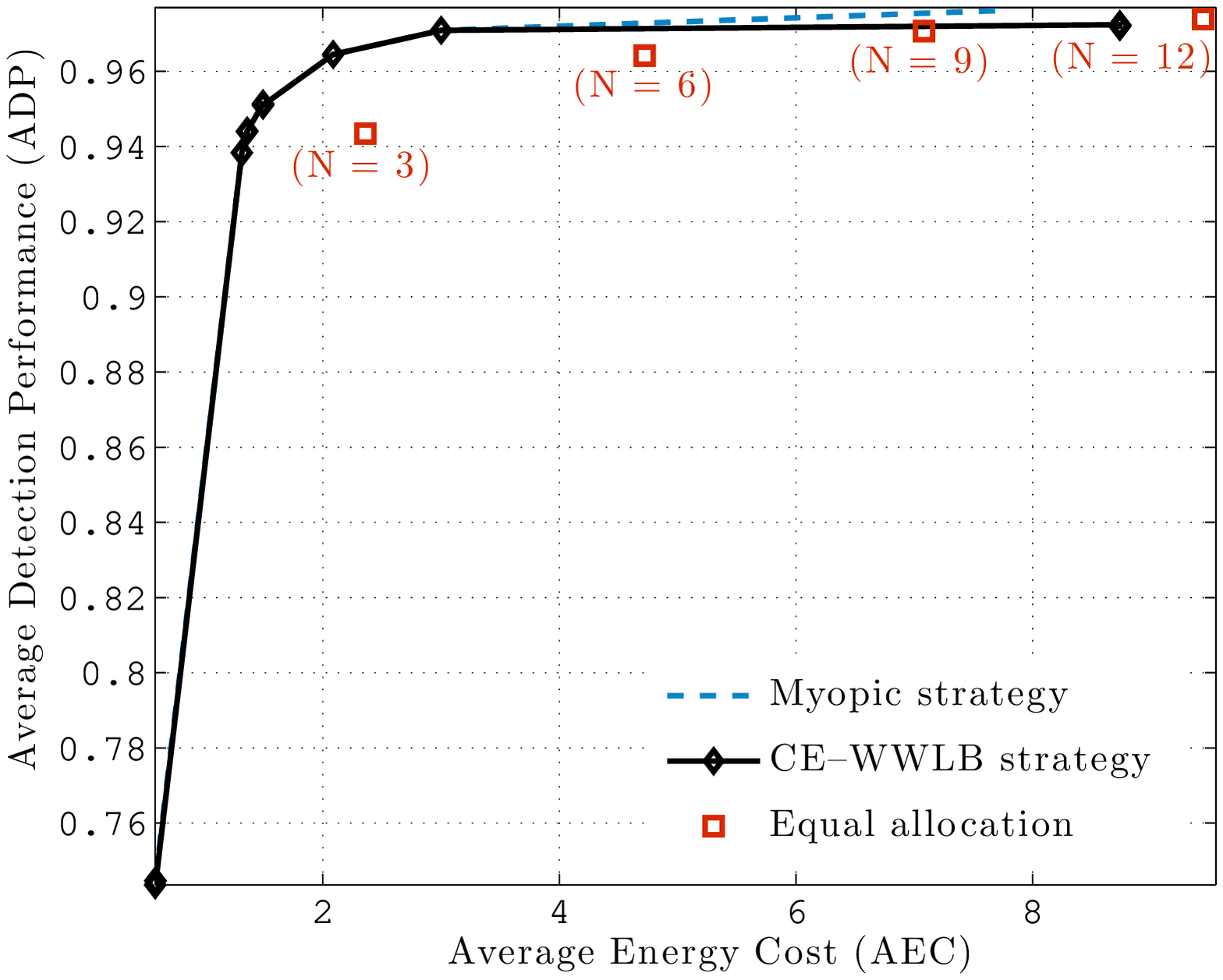}
\label{fig:trade_off_curve_DE}}
\caption{Trade--off curves for myopic ($N = 12$ samples), CE--WWLB ($N = 12$ samples) and equal allocation ($N = 3, 6, 9, 12$ samples) strategies.}\label{fig:Myopic_vs_CEWWLB_vs_EA}
\end{figure}

Fig.~\ref{fig:Myopic_vs_CEWWLB_vs_EA} illustrates the trade--off curves of the myopic and CE--WWLB strategies for $N = 12$ samples and EA for $N = 3, 6, 9, 12$. In particular, Fig.~\ref{fig:trade_off_curve_MSE} shows the AEC--AMSE curve, while Fig.~\ref{fig:trade_off_curve_DE} the AEC--ADP curve. In both cases, spending more energy leads to better MSE/detection accuracy. Furthermore, compared to EA, the two sensing strategies exhibit the same detection accuracy but lower energy consumption. We notice that the energy reduction achieved is in general identical excluding the case where detection accuracy is highly--valued. In that case, CE--WWLB spends more energy to achieve similar detection performance with the myopic strategy, as verified by Fig.~\ref{fig:Myopic_vs_CEWWLB_vs_EA}. This is due to the former strategy not using the belief state information to steer the sensor selection process, which in turn promotes a conservative selection to circumvent any worst--case scenarios. As a result, the myopic strategy exhibits $60\%$ energy gains, while CE--WWLB only $7\%$ for detection performance equal to EA's performance ($N = 12$ samples). A promising future direction is to develop sensing strategies based on on--line forms of the WWLB, that can possibly lead to larger energy gains.

\begin{figure}[h!]
\centering
\subfloat[Myopic strategy]{\includegraphics[width=0.55\linewidth]{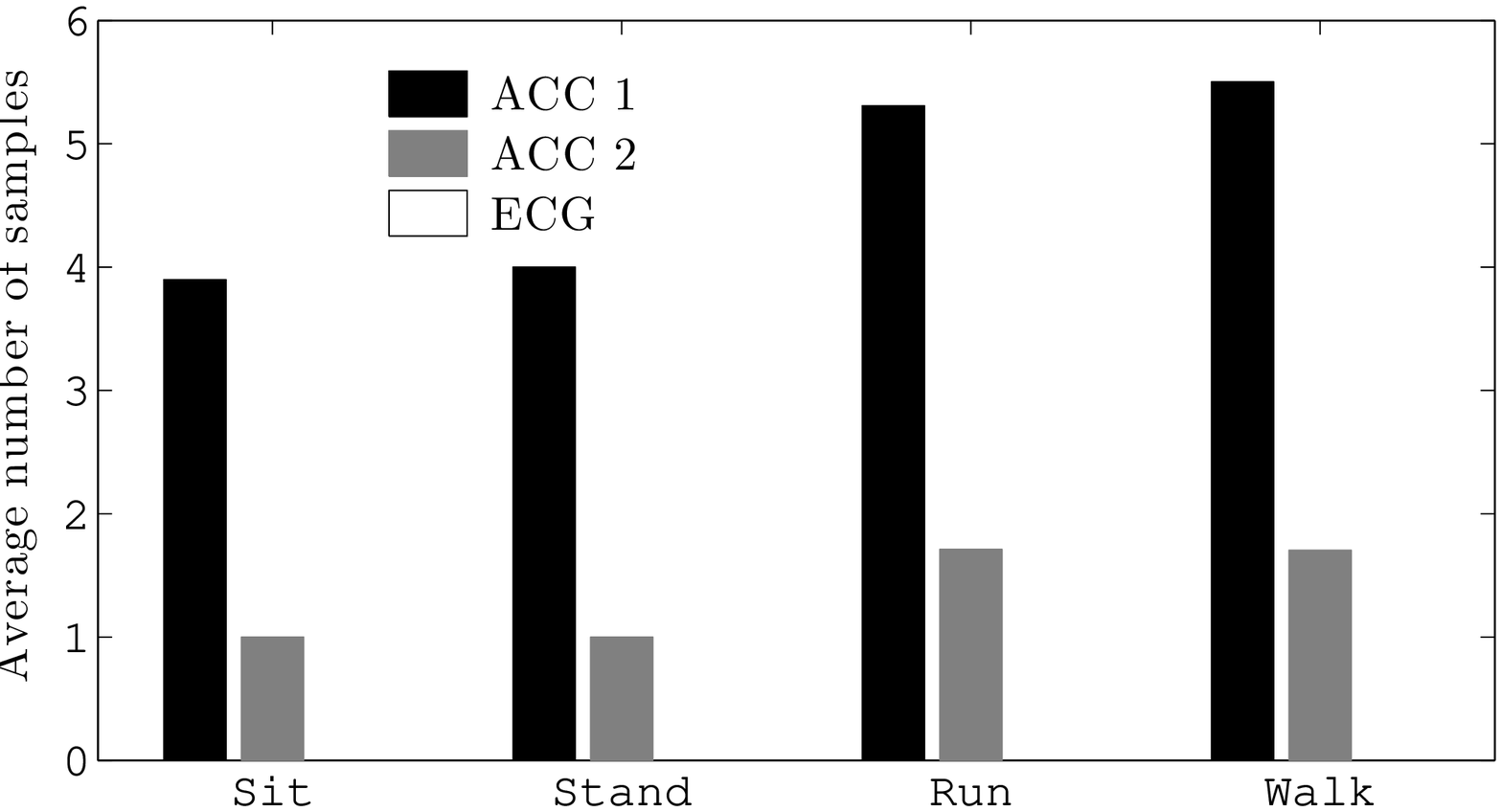}\label{fig:average_allocation_Myopic}}
\subfloat[CE--WWLB strategy]{\includegraphics[width=0.55\linewidth]{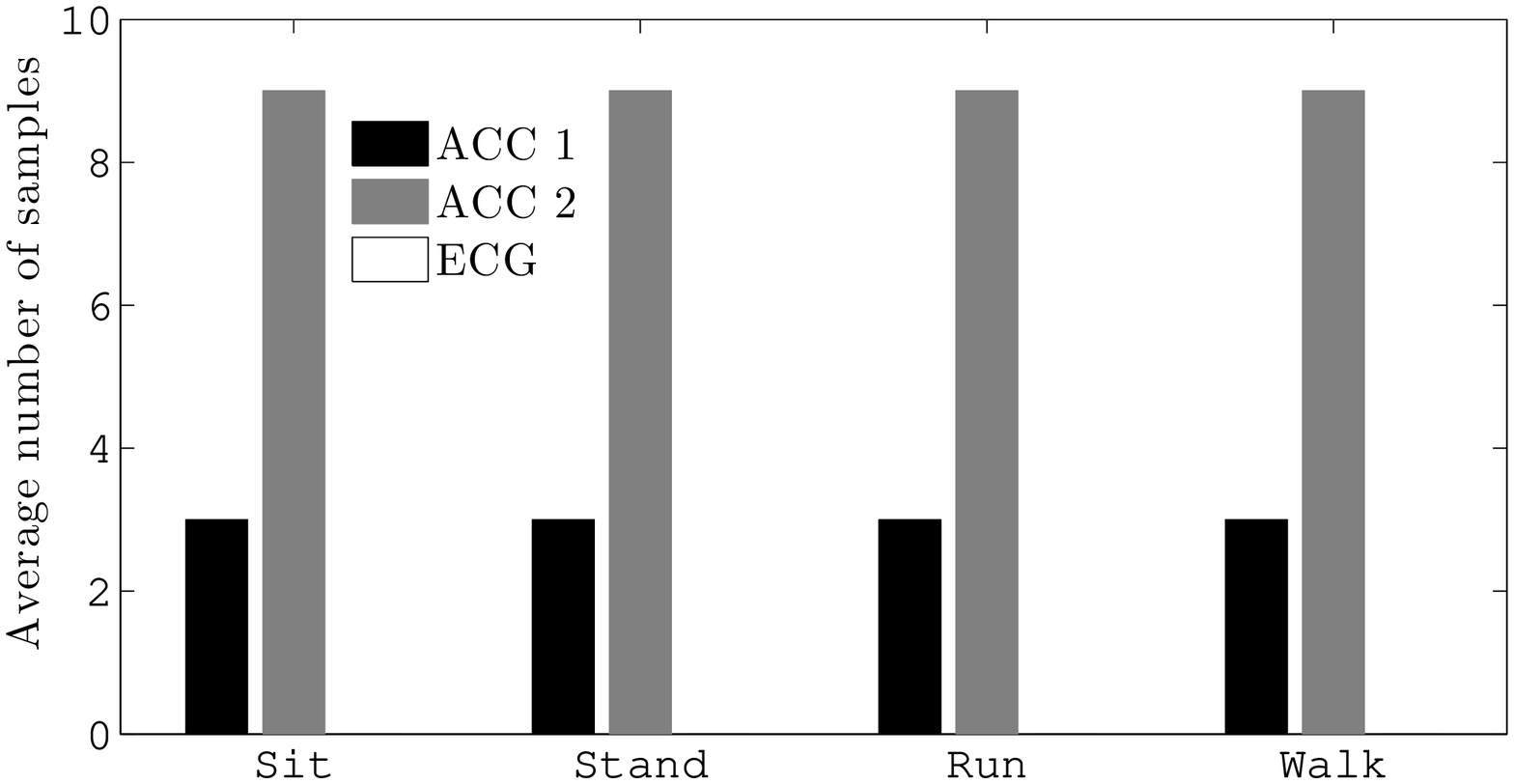}
\label{fig:average_allocation_CEWWLB}}
\caption{Samples allocation for different physical activity states for detection performance set to EA's performance ($N=12$ samples).}\label{fig:average_allocation}
\end{figure}

Finally, Fig.~\ref{fig:average_allocation} provides the average allocation of samples per sensor for the myopic (Fig.~\ref{fig:average_allocation_Myopic}) and CE--WWLB (Fig.~\ref{fig:average_allocation_CEWWLB}) strategies for the four physical activity states when their detection performance is set to EA's performance. As expected, no samples are requested from the ECG, which according to Fig.~\ref{fig:4Hypotheses}, has difficulty in distinguishing between the four physical activity states for this particular individual. On the other hand, a combination of samples from the two ACCs is used. In the myopic strategy case, the exact number depends on the physical activity of interest and on average, less than the total available samples are used. At the same time, preference is given to the first ACC. In contrast, for the CE--WWLB strategy, the exact number of samples is independent of the physical activity state since the belief state information is ignored, and preference is given to the second ACC, which is more energy--costly. Finally, neglecting belief state information and accounting for worst--case scenarios result in using all available samples.

\section{Conclusions}\label{sec:C}

In this work, we considered active state tracking of discrete--time, finite--state Markov chains observed via conditionally Gaussian measurement vectors. Our previously proposed Kalman--like estimator was employed and an optimal sensor selection strategy to optimize the trade--off between estimation performance and sensing cost was derived. Structural properties of key cost functions were also studied in conjunction with stochastic ordering. Particularly, the concavity of the cost--to--go function for non--linear POMDPs was established, which enabled us to show that the optimal policy has a threshold structure and characterize when passive sensing is optimal. Two sensing strategies with lower complexity were also presented. The proposed strategies' performance was illustrated using real data from a body sensing application, where cost--savings as high as $60\%$ were attained without significantly impairing estimation performance.

\begin{appendix}

\subsection{Proof of Theorem \ref{thm:DP_algorithm}}\label{ap:DP_proof}

The observation--control history $\mathcal{F}_{k}= \sigma \lbrace Y^{k}, U^{k-1} \rbrace$ can be iteratively rewritten as $\mathcal{F}_{k} = (\mathcal{F}_{k-1},\mathbf{y}_{k},\mathbf{u}_{k-1}), \allowbreak k = 1, 2, \dotsc, L-1, \mathcal{F}_{0} = \sigma \lbrace \mathbf{y}_{0} \rbrace,$ implying that $\mathbf{y}_{k}$ depends only on $\mathcal{F}_{k-1}$ and $\mathbf{u}_{k-1}$ since $p(\mathbf{y}_{k}|\mathcal{F}_{k-1},\mathbf{u}_{k-1}, \mathbf{y}_{0}, \allowbreak \mathbf{y}_{1}, \dotsc, \mathbf{y}_{k-1}) = p(\mathbf{y}_{k}|\mathcal{F}_{k-1},\mathbf{u}_{k-1})$. Starting form the optimal cost $J^{*}$, we exploit the conditional independence of $\mathcal{F}_{k}$ in conjunction with the iterated expectation property as follows
\begin{align}
J^{*} = &\min_{\mathbf{u}_{0},\mathbf{u}_{1},\dotsc,\mathbf{u}_{L-1}} \mathbb{E} \bigg \lbrace \sum_{k=1}^{L} g(\mathbf{y}_{k},\mathbf{u}_{k-1}) \bigg \rbrace =\min_{\mathbf{u}_{0},\mathbf{u}_{1},\dotsc,\mathbf{u}_{L-1}} \mathbb{E} \bigg \lbrace \mathbb{E} \bigg \lbrace g(\mathbf{y}_{1},\mathbf{u}_{0}) + \mathbb{E} \bigg \lbrace g(\mathbf{y}_{2},\mathbf{u}_{1}) + \dotsc \nonumber \\
&+ \mathbb{E} \bigg \lbrace g(\mathbf{y}_{L},\mathbf{u}_{L-1}) \bigg | \mathcal{F}_{L-1}, \mathbf{u}_{L-1} \bigg \rbrace \bigg | \dotsc \bigg | \mathcal{F}_{1}, \mathbf{u}_{1} \bigg \rbrace \bigg | \mathcal{F}_{0}, \mathbf{u}_{0} \bigg \rbrace \bigg \rbrace.
\end{align}

\noindent
We then use the fundamental lemma of stochastic control \cite{SpeyerChung08} to interchange expectation and minimization and get
\begin{align}
&J^{*} = \mathbb{E} \bigg \lbrace \min_{\mathbf{u}_{0}} \mathbb{E} \bigg \lbrace g(\mathbf{y}_{1},\mathbf{u}_{0}) + \min_{\mathbf{u}_{1}} \mathbb{E} \bigg \lbrace g(\mathbf{y}_{2},\mathbf{u}_{1}) + \dotsc +\min_{\mathbf{u}_{L-1}} \mathbb{E} \bigg \lbrace g(\mathbf{y}_{L},\mathbf{u}_{L-1}) \bigg | \mathcal{F}_{L-1}, \mathbf{u}_{L-1} \bigg \rbrace \bigg | \dotsc \bigg | \mathcal{F}_{1}, \mathbf{u}_{1}  \bigg \rbrace \bigg | \mathcal{F}_{0}, \mathbf{u}_{0} \bigg \rbrace \bigg \rbrace.
\end{align}

\noindent
Employing the principle of optimality \cite{BertsekasDPOC05} that applies to dynamic decision problems with sum cost functions, we get
\begin{align}\label{eq:DP_recurrence_initial}
J_{L}(\mathcal{F}_{L-1})& = \min_{\mathbf{u}_{L-1} \in \mathcal{U}} \bigg [ \underset{\mathbf{y}_{L}}{\mathbb{E}} \big \lbrace g(\mathbf{y}_{L},\mathbf{u}_{L-1}) \big | \mathcal{F}_{L-1}, \mathbf{u}_{L-1} \big \rbrace \bigg ], \nonumber \\
J_{L-1}(\mathcal{F}_{L-2})& = \min_{\mathbf{u}_{L-2} \in \mathcal{U}} \bigg [ \underset{\mathbf{y}_{L-1}}{\mathbb{E}} \big \lbrace g(\mathbf{y}_{L-1},\mathbf{u}_{L-2}) \nonumber \\
&+ J_{L}(\mathcal{F}_{L-2},\mathbf{y}_{L-1},\mathbf{u}_{L-2}) \big | \mathcal{F}_{L-2}, \mathbf{u}_{L-2} \big \rbrace \bigg ], \\
&\vdots \nonumber \\
J_{1}(\mathcal{F}_{0})& = \min_{\mathbf{u}_{0} \in \mathcal{U}} \bigg [ \underset{\mathbf{y}_{1}}{\mathbb{E}} \big \lbrace g(\mathbf{y}_{1},\mathbf{u}_{0}) + J_{2}(\mathcal{F}_{0},\mathbf{y}_{1},\mathbf{u}_{0}) \big | \mathcal{F}_{0}, \mathbf{u}_{0} \big \rbrace \bigg ].\nonumber
\end{align}

Since the dimension of $\mathcal{F}_{k-1}$ increases at each time slot $k - 1$ with the addition of a new observation and control, we use $\mathbf{p}_{k|k-1}$ as a sufficient statistic for control purposes \cite{ZoisTSP14}. Then, we rewrite (\ref{eq:DP_recurrence_initial}) as a function of $\mathbf{p}_{k|k-1}$ by separately computing each term inside the minimization in (\ref{eq:DP_recurrence_initial}). Specifically, for the first term, we have
\begin{align}\label{eq:first_term}
\underset{\mathbf{y}_{k}}{\mathbb{E}} \big \lbrace g(\mathbf{y}_{k},\mathbf{u}_{k-1}) \big | \mathcal{F}_{k-1}, \mathbf{u}_{k-1} \big \rbrace& = (1-\lambda) \underset{\mathbf{y}_{k}}{\mathbb{E}} \big \lbrace \tr(\mathbf{\Sigma}_{k|k}(\mathbf{y}_{k},\mathbf{u}_{k-1})) \big | \mathcal{F}_{k-1}, \mathbf{u}_{k-1} \big \rbrace + \lambda \underset{\mathbf{y}_{k}}{\mathbb{E}} \big \lbrace c(\mathbf{u}_{k-1}) \big | \mathcal{F}_{k-1}, \mathbf{u}_{k-1} \big \rbrace \nonumber \\
&\overset{(a)}{=} (1-\lambda) \sum_{i=1}^{n} p_{k|k-1}^{i} (1 - \tr(\mathbf{G}_{k}^{T} \mathbf{G}_{k} \mathbf{Q}_{i}^{\mathbf{u}_{k-1}}) - \norm{\mathbf{p}_{k|k-1} + \mathbf{G}_{k}(\mathbf{m}_{i}^{\mathbf{u}_{k-1}} - \mathbf{y}_{k|k-1})}^{2}) \nonumber \\
& + \lambda \int p(\mathbf{y} | \mathcal{F}_{k-1}, \mathbf{u}_{k-1})c(\mathbf{u}_{k-1}) d \mathbf{y} = (1-\lambda)\mathbf{p}_{k|k-1}^{T} \mathbf{h}(\mathbf{p}_{k|k-1}, \mathbf{u}_{k-1}) + \lambda c(\mathbf{u}_{k-1}) \nonumber \\
&\doteq \ell(\mathbf{p}_{k|k-1},\mathbf{u}_{k-1}),
\end{align}

\noindent
where $(a)$ the first term has been derived in \cite{ZoisTSP14} and the second term is by the definition of conditional expectation, and $\mathbf{h}(\mathbf{p}_{k|k-1},\mathbf{u}_{k-1}) = [h(\mathbf{e}_{1},\mathbf{p}_{k|k-1},\mathbf{u}_{k-1}), \dotsc, h(\mathbf{e}_{n},\mathbf{p}_{k|k-1},\mathbf{u}_{k-1})]^{T}$ is a $n$--dimensional vector with $h(\mathbf{e}_{i},\mathbf{p}_{k|k-1},\mathbf{u}_{k-1}) = 1 - \tr(\mathbf{G}_{k}^{T} \mathbf{G}_{k} \mathbf{Q}_{i}^{\mathbf{u}_{k-1}}) - \norm{\mathbf{p}_{k|k-1} + \mathbf{G}_{k}(\mathbf{m}_{i}^{\mathbf{u}_{k-1}} - \mathbf{y}_{k|k-1})}^{2}$. The second term in (\ref{eq:DP_recurrence_initial}) can be computed as
\begin{align}\label{eq:second_term}
\underset{\mathbf{y}_{k}}{\mathbb{E}} \big \lbrace J_{k+1}(\mathcal{F}_{k-1},\mathbf{y}_{k},\mathbf{u}_{k-1}) \big | \mathcal{F}_{k-1}, \mathbf{u}_{k-1} \big \rbrace& = \underset{\mathbf{y}_{k}}{\mathbb{E}} \big \lbrace \overline{J}_{k+1}(\Phi_{k}(\mathbf{p}_{k-1},\mathbf{y}_{k},\mathbf{u}_{k-1})) \big | \mathbf{p}_{k-1}, \mathbf{u}_{k-1} \big \rbrace \nonumber \\
&= \int p(\mathbf{y}|\mathbf{p}_{k|k-1},\mathbf{u}_{k|k-1}) \overline{J}_{k+1}(\Phi_{k}(\mathbf{p}_{k-1},\mathbf{y},\mathbf{u}_{k-1})) d\mathbf{y} \nonumber \\
&= \int \mathbf{1}_{n}^{T} \mathbf{r}(\mathbf{y},\mathbf{u}_{k-1})\mathbf{p}_{k|k-1} \overline{J}_{k+1}\bigg ( \frac{\mathbf{P} \mathbf{r}(\mathbf{y},\mathbf{u}_{k-1}) \mathbf{p}_{k|k-1}}{\mathbf{1}_{n}^{T}\mathbf{r}(\mathbf{y},\mathbf{u}_{k-1}) \mathbf{p}_{k|k-1}} \bigg )d \mathbf{y},
\end{align}

\noindent
where we have used the fact that $\mathbf{p}_{k|k-1}$ is a sufficient statistic of $\mathcal{F}_{k-1}$, $\mathbf{u}_{k-1} = \eta_{k-1}(\mathcal{F}_{k-1})$ and the update rule in (\ref{eq:update_rule}). Substituting (\ref{eq:first_term}) -- (\ref{eq:second_term}) back to (\ref{eq:DP_recurrence_initial}), we get
\begin{align}
\overline{J}_{L}(\mathbf{p}_{L|L-1})& = \min_{\mathbf{u}_{L-1} \in \mathcal{U}} \big [ \ell(\mathbf{p}_{L|L-1},\mathbf{u}_{L-1}) \big ] \nonumber \\
\overline{J}_{L-1}(\mathbf{p}_{L-1|L-2})& = \min_{\mathbf{u}_{L-2} \in \mathcal{U}} \bigg [ \ell(\mathbf{p}_{L-1|L-2},\mathbf{u}_{L-2}) + \int \mathbf{1}_{n}^{T} \mathbf{r}(\mathbf{y},\mathbf{u}_{L-2})\mathbf{p}_{L-1|L-2}\overline{J}_{L}\bigg ( \frac{\mathbf{P} \mathbf{r}(\mathbf{y},\mathbf{u}_{L-2}) \mathbf{p}_{L-1|L-2}}{\mathbf{1}_{n}^{T}\mathbf{r}(\mathbf{y},\mathbf{u}_{L-2}) \mathbf{p}_{L-1|L-2}} \bigg ) d \mathbf{y} \bigg ], \\
&\vdots \nonumber \\
\overline{J}_{1}(\mathbf{p}_{1|0})& = \min_{\mathbf{u}_{0} \in \mathcal{U}} \bigg [ \ell(\mathbf{p}_{1|0},\mathbf{u}_{0}) + \int \mathbf{1}_{n}^{T} \mathbf{r}(\mathbf{y},\mathbf{u}_{0})\mathbf{p}_{1|0}\overline{J}_{2}\bigg ( \frac{\mathbf{P} \mathbf{r}(\mathbf{y},\mathbf{u}_{0}) \mathbf{p}_{1|0}}{\mathbf{1}_{n}^{T}\mathbf{r}(\mathbf{y},\mathbf{u}_{0}) \mathbf{p}_{1|0}} \bigg ) d \mathbf{y} \bigg ]. \nonumber
\end{align}

\subsection{Proof of Lemma \ref{lem:current_cost_rw}}\label{ap:current_cost_rw}

The current cost of selecting control input $\mathbf{u}_{k-1}$ consists of two parts, the estimation error part and the sensing cost part
\begin{equation}\label{eq:current_cost_initial_fm}
\ell(\mathbf{p}_{k|k-1},\mathbf{u}_{k-1}) = (1-\lambda)\mathbf{p}_{k|k-1}^{T} \mathbf{h}(\mathbf{p}_{k|k-1}, \mathbf{u}_{k-1}) + \lambda c(\mathbf{u}_{k-1}).
\end{equation}

\noindent
We simplify the former part as follows
\begin{align}\label{eq:cc_estimationP_00}
&\mathbf{p}_{k|k-1}^{T} \mathbf{h}(\mathbf{p}_{k|k-1}) = \sum_{i=1}^{n} p_{k|k-1}^{i} - \sum_{i=1}^{n} p_{k|k-1}^{i} \tr{(\mathbf{G}_{k} \mathbf{G}_{k}^{T} \mathbf{Q}_{i}^{\mathbf{u}_{k-1}})} -\sum_{i=1}^{n} p_{k|k-1}^{i} \norm{\mathbf{p}_{k|k-1} + \mathbf{G}_{k}(\mathbf{m}_{i}^{\mathbf{u}_{k-1}} - \mathbf{y}_{k|k-1})}^2.
\end{align}

\noindent
At this point, we compute each term in (\ref{eq:cc_estimationP_00}) separately. Clearly, the first term $\sum_{i=1}^{n} p_{k|k-1}^{i}$ equals $1$; for the second term, we exploit the linearity of the trace operator as follows
\begin{align}\label{eq:cc_estimationP_01}
\sum_{i=1}^{n} p_{k|k-1}^{i} \tr{(\mathbf{G}_{k} \mathbf{G}_{k}^{T} \mathbf{Q}_{i}^{\mathbf{u}_{k-1}})}& = \tr{(\mathbf{G}_{k} \mathbf{G}_{k}^{T} \sum_{i=1}^{n} p_{k|k-1}^{i}\mathbf{Q}_{i}^{\mathbf{u}_{k-1}})} = \tr{(\mathbf{G}_{k} \mathbf{G}_{k}^{T} \widetilde{\mathbf{Q}}_{k})},
\end{align}

\noindent
where in the last step, we have used the definition of $\widetilde{\mathbf{Q}}_{k}$ in Theorem \ref{thm:approx_MMSE_estimate}. For the third term, we have
\begin{align}\label{eq:cc_estimationP_02}
&\sum_{i=1}^{n} p_{k|k-1}^{i} \norm{\mathbf{p}_{k|k-1} + \mathbf{G}_{k}(\mathbf{m}_{i}^{\mathbf{u}_{k-1}} - \mathbf{y}_{k|k-1})}^2 = \tr ( \mathbf{p}_{k|k-1} \mathbf{p}_{k|k-1}^{T} + \sum_{i=1}^{n} p_{k|k-1}^{i} \mathbf{p}_{k|k-1}(\mathbf{m}_{i}^{\mathbf{u}_{k-1}} - \mathbf{y}_{k|k-1})^{T} \mathbf{G}_{k}^{T} \nonumber \\
&+ \sum_{i=1}^{n} p_{k|k-1}^{i} \mathbf{G}_{k} (\mathbf{m}_{i}^{\mathbf{u}_{k-1}} - \mathbf{y}_{k|k-1}) \mathbf{p}_{k|k-1}^{T} + \sum_{i=1}^{n} p_{k|k-1}^{i} \mathbf{G}_{k} (\mathbf{m}_{i}^{\mathbf{u}_{k-1}} - \mathbf{y}_{k|k-1})(\mathbf{m}_{i}^{\mathbf{u}_{k-1}} - \mathbf{y}_{k|k-1})^{T} \mathbf{G}_{k}^{T})
\end{align}

\noindent
For the second term inside the trace operator above, we have
\begin{align}\label{eq:cc_estimationP_03}
\sum_{i=1}^{n} p_{k|k-1}^{i} \mathbf{p}_{k|k-1}(\mathbf{m}_{i}^{\mathbf{u}_{k-1}} - \mathbf{y}_{k|k-1})^{T} \mathbf{G}_{k}^{T}& = \mathbf{p}_{k|k-1} \big ( \sum_{i}^{n} p_{k|k-1}^{i} \mathbf{m}_{i}^{\mathbf{u}_{k-1},T} - \sum_{i=1}^{n} p_{k|k-1}^{i}  \mathbf{y}_{k|k-1}^{T} \big ) \mathbf{G}_{k}^{T} \nonumber \\
&= \mathbf{p}_{k|k-1} (( \mathcal{M}(\mathbf{u}_{k-1}) \mathbf{p}_{k|k-1})^{T} - \mathbf{y}_{k|k-1}^{T}) \mathbf{G}_{k}^{T} = 0.
\end{align}

\noindent
Similarly, the third term inside the trace operator is equal to zero. Lastly, for the fourth term, we get
\begin{align}\label{eq:cc_estimationP_04}
\sum_{i=1}^{n} p_{k|k-1}^{i} \mathbf{G}_{k} (\mathbf{m}_{i}^{\mathbf{u}_{k-1}} - \mathbf{y}_{k|k-1}) (\mathbf{m}_{i}^{\mathbf{u}_{k-1}} - \mathbf{y}_{k|k-1})^{T} \mathbf{G}_{k}^{T}& = \mathbf{G}_{k} ( \mathcal{M}(\mathbf{u}_{k-1}) \diag(\mathbf{p}_{k|k-1}) \mathcal{M}^{T}(\mathbf{u}_{k-1}) \nonumber \\
&- \mathbf{y}_{k|k-1} \mathbf{y}_{k|k-1}^{T}) \mathbf{G}_{k}^{T}.
\end{align}

\noindent
Substituting (\ref{eq:cc_estimationP_01}) and (\ref{eq:cc_estimationP_04}) back to (\ref{eq:cc_estimationP_00}), we get
\begin{align}\label{eq:cc_estimationP_05}
\ell(\mathbf{p}_{k|k-1},\mathbf{u}_{k-1})& = 1 - \tr( \mathbf{p}_{k|k-1} \mathbf{p}_{k|k-1}^{T} + \mathbf{\Sigma}_{k|k-1} \mathcal{M}^{T}(\mathbf{u}_{k-1}) \mathbf{G}_{k}^{T}) \nonumber \\
& = \tr(\diag(\mathbf{p}_{k|k-1}) - \mathbf{p}_{k|k-1} \mathbf{p}_{k|k-1}^{T} - \mathbf{\Sigma}_{k|k-1} \mathcal{M}^{T}(\mathbf{u}_{k-1})\mathbf{G}_{k}^{T}) \nonumber \\
& = \tr{( (\mathbf{I} - \mathbf{G}_{k} \mathcal{M}(\mathbf{u}_{k-1} ) ) \mathbf{\Sigma}_{k|k-1} )},
\end{align}

\noindent
where we have exploited that $\tr(A^{T}) = \tr(A)$. Substituting (\ref{eq:cc_estimationP_05}) back in (\ref{eq:current_cost_initial_fm}) concludes the proof.

\subsection{Proof of Lemma \ref{lem:concave_current_cost}}\label{ap:concave_current_cost}

For clarity, we drop the dependence on time. We focus on discriminating between two states, $\mathbf{e}_{1}$ and $\mathbf{e}_{2}$, hence the predicted belief state is of the form $\mathbf{p} = [p,1-p]^{T}$. Thus, after some manipulations, the current cost term becomes
\begin{align}\label{eq:concave_current_cost_00}
\ell(p,\mathbf{u})& = (1-\lambda)(2f(p) - 2f^{2}(p) \tr(\mathbf{L} \mathcal{M}^{T}(\mathbf{u}) (f(p)\mathcal{M}(\mathbf{u}) \mathbf{L} \nonumber \\
& \times \mathcal{M}^{T}(\mathbf{u}) + p \mathbf{Q}_{1}^{\mathbf{u}} + (1-p) \mathbf{Q}_{2}^{\mathbf{u}} )^{-1} \mathcal{M}(\mathbf{u}))) + p \mathbf{Q}_{1}^{\mathbf{u}} + (1-p) \mathbf{Q}_{2}^{\mathbf{u}} )^{-1} \mathcal{M}(\mathbf{u}))) + \lambda c(\mathbf{u}),
\end{align}

\noindent
where $f(p) = p(1-p)$ and $\mathbf{L} = \left[ \begin{smallmatrix} 1&-1\\ -1&1 \end{smallmatrix} \right]$. The function $f(p)$ is a concave function of $p$. Since we have scalar measurements, (\ref{eq:concave_current_cost_00}) becomes
\begin{align}\label{eq:concave_current_cost_01}
\ell(p,\mathbf{u})& = (1-\lambda)(2f(p) - 2f^{2}(p) \tr(\mathbf{L} [m_{1}^{\mathbf{u}},m_{2}^{\mathbf{u}}]^{T} (f(p)[m_{1}^{\mathbf{u}},m_{2}^{\mathbf{u}}] \mathbf{L} [m_{1}^{\mathbf{u}},m_{2}^{\mathbf{u}}]^{T}+ p \sigma_{1,\mathbf{u}}^{2} + (1-p) \sigma_{2,\mathbf{u}}^{2})^{-1} \nonumber \\
&[m_{1}^{\mathbf{u}},m_{2}^{\mathbf{u}}] )) + \lambda c(\mathbf{u}) = ( 1-\lambda) \bigg ( 2f(p) - \frac{2a_{12}(\mathbf{u}) f^{2}(p)}{a_{12}(\mathbf{u})f(p) + \sigma_{1,\mathbf{u}}^{2}p + \sigma_{2,\mathbf{u}}^{2}(1-p)} \bigg ) + \lambda c(\mathbf{u}),
\end{align}

\noindent
where $a_{12}(\mathbf{u}) = (m_{1}^{\mathbf{u}}-m_{2}^{\mathbf{u}})^2 \geqslant 0$. In order to characterize Eq.~(\ref{eq:concave_current_cost_01}), we distinguish between the cases: 1) $m_{1}^{\mathbf{u}} = m_{2}^{\mathbf{u}}$ and $\sigma_{1,\mathbf{u}}^{2} = \sigma_{2,\mathbf{u}}^{2}$, $\mathbf{u} \in \mathcal{U}$ (Case I), 2) $m_{1}^{\mathbf{u}} = m_{2}^{\mathbf{u}}$ and $\sigma_{1,\mathbf{u}}^{2} \neq \sigma_{2,\mathbf{u}}^{2}$, $\mathbf{u} \in \mathcal{U}$ (Case II), 3) $m_{1}^{\mathbf{u}} \neq m_{2}^{\mathbf{u}}$ and $\sigma_{1,\mathbf{u}}^{2} = \sigma_{2,\mathbf{u}}^{2}$, $\mathbf{u} \in \mathcal{U}$ (Case III), and 4) $m_{1}^{\mathbf{u}} \neq m_{2}^{\mathbf{u}}$ and $\sigma_{1,\mathbf{u}}^{2} \neq \sigma_{2,\mathbf{u}}^{2}$, $\mathbf{u} \in \mathcal{U}$ (Case IV).

For Cases I and II, $a_{12}(\mathbf{u}) = 0$, and thus Eq.~(\ref{eq:concave_current_cost_01}) becomes $\ell(p,\mathbf{u}) = 2(1-\lambda)f(p) + \lambda c(\mathbf{u})$. The latter expression is a concave function of $p$ and depends on the control input $\mathbf{u}$ through the sensing cost $c(\mathbf{u})$. For Case III, Eq.~(\ref{eq:concave_current_cost_01}) becomes
\begin{equation}\label{eq:concave_current_cost_03}
\ell(p,\mathbf{u}) =  (1-\lambda)\frac{2\sigma_{\mathbf{u}}^{2}f(p)}{a_{12}(\mathbf{u})f(p) + \sigma_{\mathbf{u}}^{2}} + \lambda c(\mathbf{u}),
\end{equation}

\noindent
where $a_{12}(\mathbf{u}) > 0$ and $\sigma_{\mathbf{u}}^{2} = \sigma_{1,\mathbf{u}}^{2} = \sigma_{2,\mathbf{u}}^{2}$. Its second derivative with respect to $p$ has the form
\begin{equation}\label{eq:concave_current_cost_04}
\ell''(p,\mathbf{u}) = -\frac{4\sigma_{u}^{4} (a_{12}(\mathbf{u})(3p(p-1) + 1) + \sigma_{\mathbf{u}}^{2})}{(a_{12}(u)f(p) + \sigma_{\mathbf{u}}^{2})^3} < 0,
\end{equation}

\noindent
where the last inequality holds $\forall p \in [0,1]$ since $a_{12}(u) > 0$, $f(p) \geqslant 0$ and $3p(p-1) + 1 > 0, \forall p \in [0,1]$. As a result, the current cost in (\ref{eq:concave_current_cost_03}) is also a concave function of $p$. Finally, for Case IV, the current cost in (\ref{eq:concave_current_cost_01}) takes the form
\begin{equation}\label{eq:concave_current_cost_05}
\ell(p,\mathbf{u}) =  (1-\lambda)\frac{2f(p)(\sigma_{1,u}^{2}p + \sigma_{2,u}^{2}(1-p))}{a_{12}(u)f(p) + \sigma_{1,u}^{2}p + \sigma_{2,u}^{2}(1-p)} + \lambda c(\mathbf{u})
\end{equation}

\noindent
where $a_{12}(u) > 0$. The second derivative with respect to $p$ is
\begin{equation}\label{eq:concave_current_cost_06}
\ell''(p,\mathbf{u}) = -\frac{\alpha_{p,a_{12}(u),\sigma_{1,u}^2} + \beta_{p,a_{12}(u),\sigma_{2,u}^2} +  \gamma_{p,\sigma_{1,u}^2,\sigma_{2,u}^2} }{(\sigma_{1,u}^{2}p + \sigma_{2,u}^{2}(1-p) + a_{12}(u)f(p))^3},
\end{equation}

\noindent
where $\alpha_{p,a_{12}(u),\sigma_{1,u}^2} = \sigma_{1,u}^{4}(1 + a_{12}(u)\sigma_{1,u}^{2})p^{3}$, $\beta_{p,a_{12}(u),\sigma_{2,u}^2} = \sigma_{2,u}^{4} (\sigma_{2,u}^2 + a_{12}(u))(1-p)^3$ and $\gamma_{p,\sigma_{1,u}^2,\sigma_{2,u}^2} = \sigma_{1,u}^2 \sigma_{2,u}^2 f(p) (\sigma_{1,u}^{2}p + 3\sigma_{2,u}^{2}(1-p))$. Each of the latter terms is greater than or equal to zero yielding that the numerator in (\ref{eq:concave_current_cost_06}) is greater than zero. The denominator in (\ref{eq:concave_current_cost_06}) is also greater than zero. Thus, the second derivative given in (\ref{eq:concave_current_cost_06}) is negative $\forall p \in [0,1]$ and therefore, the current cost in (\ref{eq:concave_current_cost_05}) constitutes a concave function of $p$.

\subsection{Proof of Theorem \ref{thm:concave_cost2go}}\label{ap:concave_cost2go}

We prove the concavity of the cost--to--go function $\overline{J}_{k}(\mathbf{p}_{k|k-1})$ by induction. At time step $L$, it is clear that $\overline{J}_{L}(\mathbf{p}_{L|L-1})$ is a concave function since according to Lemma \ref{lem:concave_current_cost}, for each $\mathbf{u}_{L} \in \mathcal{U}, \ell(\mathbf{p}_{L|L-1},\mathbf{u}_{L})$ is a concave function and the pointwise minimum of concave functions is also concave.

Next, we assume that $\overline{J}_{k+1}(\mathbf{p}_{k+1|k})$ is concave, and to prove the concavity of $\overline{J}_{k}(\mathbf{p}_{k|k-1})$, we only need to show that $\int \mathbf{1}_{n}^{T} \mathbf{r}(\mathbf{y},\mathbf{u}_{k-1}) \mathbf{p}_{k|k-1} \overline{J}_{k+1}(\boldsymbol \Phi (\mathbf{p}_{k|k-1},\mathbf{y},\mathbf{u}_{k-1})) d \mathbf{y}$, where $\boldsymbol \Phi (\cdot)$ denotes the associated update rule, is also a concave function for all $\mathbf{u}_{k-1} \in \mathcal{U}$. Let $\mathbf{v}$ and $\mathbf{w}$ two predicted belief state vectors. For any $\alpha, 0 \leqslant \alpha \leqslant 1$, we have
\begin{align}
&\alpha \int \mathbf{1}_{n}^{T} \mathbf{r}(\mathbf{y},\mathbf{u}_{k-1}) \mathbf{v}\overline{J}_{k+1}(\boldsymbol \Phi (\mathbf{v},\mathbf{y},\mathbf{u}_{k-1})) d \mathbf{y} +  
(1-\alpha) \int \mathbf{1}_{n}^{T} \mathbf{r}(\mathbf{y},\mathbf{u}_{k-1}) \mathbf{w}\overline{J}_{k+1}(\boldsymbol \Phi (\mathbf{w},\mathbf{y},\mathbf{u}_{k-1})) d \mathbf{y} = \nonumber \\
&\int (\alpha \mathbf{1}_{n}^{T} \mathbf{r}(\mathbf{y},\mathbf{u}_{k-1}) \mathbf{v} + (1-\alpha) \mathbf{1}_{n}^{T} \mathbf{r}(\mathbf{y},\mathbf{u}_{k-1}) \mathbf{w} ) \bigg [ \frac{\alpha \mathbf{1}_{n}^{T} \mathbf{r}(\mathbf{y},\mathbf{u}_{k-1}) \mathbf{v} \overline{J}_{k+1}(\boldsymbol \Phi (\mathbf{v},\mathbf{y},\mathbf{u}_{k-1}))}{\alpha \mathbf{1}_{n}^{T} \mathbf{r}(\mathbf{y},\mathbf{u}_{k-1}) \mathbf{v} + (1-\alpha) \mathbf{1}_{n}^{T} \mathbf{r}(\mathbf{y},\mathbf{u}_{k-1})\mathbf{w} } \nonumber \\
& + \frac{\alpha \mathbf{1}_{n}^{T} \mathbf{r}(\mathbf{y},\mathbf{u}_{k-1}) \mathbf{w} \overline{J}_{k+1}(\boldsymbol \Phi (\mathbf{v},\mathbf{y},\mathbf{u}_{k-1}))}{\alpha \mathbf{1}_{n}^{T} \mathbf{r}(\mathbf{y},\mathbf{u}_{k-1}) \mathbf{v} + (1-\alpha) \mathbf{1}_{n}^{T} \mathbf{r}(\mathbf{y},\mathbf{u}_{k-1})\mathbf{w} }  \bigg ] d \mathbf{y} \leqslant \int (\alpha \mathbf{1}_{n}^{T} \mathbf{r}(\mathbf{y},\mathbf{u}_{k-1}) \mathbf{v} + (1-\alpha) \mathbf{1}_{n}^{T} \mathbf{r}(\mathbf{y},\mathbf{u}_{k-1}) \mathbf{w} ) \nonumber \\
&\times \overline{J}_{k+1} \bigg ( \frac{\alpha \mathbf{1}_{n}^{T} \mathbf{r}(\mathbf{y},\mathbf{u}_{k-1}) \mathbf{v} \boldsymbol \Phi (\mathbf{v},\mathbf{y},\mathbf{u}_{k-1})}{\alpha \mathbf{1}_{n}^{T} \mathbf{r}(\mathbf{y},\mathbf{u}_{k-1}) \mathbf{v} +(1-\alpha) \mathbf{1}_{n}^{T} \mathbf{r}(\mathbf{y},\mathbf{u}_{k-1}) \mathbf{w}} + \frac{(1-\alpha) \mathbf{1}_{n}^{T} \mathbf{r}(\mathbf{y},\mathbf{u}_{k-1}) \mathbf{w} \boldsymbol \Phi (\mathbf{w},\mathbf{y},\mathbf{u}_{k-1})}{\alpha \mathbf{1}_{n}^{T} \mathbf{r}(\mathbf{y},\mathbf{u}_{k-1}) \mathbf{v} +(1-\alpha) \mathbf{1}_{n}^{T} \mathbf{r}(\mathbf{y},\mathbf{u}_{k-1}) \mathbf{w}} \bigg) d \mathbf{y} = \nonumber \\
&\int (\alpha \mathbf{1}_{n}^{T} \mathbf{r}(\mathbf{y},\mathbf{u}_{k-1}) \mathbf{v} +(1-\alpha) \mathbf{1}_{n}^{T} \mathbf{r}(\mathbf{y},\mathbf{u}_{k-1}) \mathbf{w})\overline{J}_{k+1} (\boldsymbol \Phi(\alpha \mathbf{v} + (1-\alpha)\mathbf{w}, \mathbf{y}, \mathbf{u}_{k-1}) )d\mathbf{y},
\end{align}

\noindent
where the inequality comes from the induction hypothesis and the last step implies that for all $\mathbf{u}_{k-1} \in \mathcal{U}$, the function  $\int \mathbf{1}_{n}^{T} \mathbf{r}(\mathbf{y},\mathbf{u}_{k-1}) \mathbf{p}_{k|k-1} \overline{J}_{k+1}(\boldsymbol \Phi (\mathbf{p}_{k|k-1},\mathbf{y},\mathbf{u}_{k-1})) \allowbreak d \mathbf{y}$ is concave. Last but not least, $\overline{J}_{k}(\mathbf{p}_{k|k-1})$ constitutes the minimum of concave functions and thus, it is also concave.

\subsection{Proof of Corollary \ref{cor:diff_means_same_vars}}\label{ap:diff_means_same_vars}

We start from (\ref{eq:diff_means_same_vars}) and consider two cases: 1) $c(\mathbf{u}^{i}) = c, \forall \mathbf{u}^{i} \in \mathcal{U}$ and $c$ constant, 2) $c(\mathbf{u}^{i}) < c(\mathbf{u}^{j}), \mathbf{u}^{i}, \mathbf{u}^{j} \in \mathcal{U}$ with $i \neq j$. For the first case, we see that
\begin{align}
\ell(p,\mathbf{u}^{i})& \geqslant \ell(p,\mathbf{u}^{j}) \Rightarrow \nonumber \\
(1-\lambda)\frac{2\sigma_{\mathbf{u}}^2 f(p)}{a_{12}(\mathbf{u}^{i}) f(p) + \sigma_{\mathbf{u}}^2}  + \lambda c(\mathbf{u}^{i}) &\geqslant (1-\lambda) \frac{2\sigma_{\mathbf{u}}^2 f(p)}{a_{12}(\mathbf{u}^{j}) f(p) + \sigma_{\mathbf{u}}^2} + \lambda c(\mathbf{u}^{j}) \Rightarrow \nonumber \\
\frac{1}{a_{12}(\mathbf{u}^{i}) f(p) + \sigma_{\mathbf{u}}^2} &\geqslant \frac{1}{a_{12}(\mathbf{u}^{j}) f(p) + \sigma_{\mathbf{u}}^2} \Rightarrow \nonumber \\
a_{12}(\mathbf{u}^{i}) &\leqslant a_{12}(\mathbf{u}^{j}),
\end{align}

\noindent
which implies that ordering of controls can be achieved based on $a_{12}(\mathbf{u}) = (m_{1}^{\mathbf{u}} - m_{2}^{\mathbf{u}})^2$. For the second case, we assume that for controls $\mathbf{u}^{i}, \mathbf{u}^{j} \in \mathcal{U}, i \neq j$, $a_{12}(\mathbf{u}^{i}) > a_{12}(\mathbf{u}^{j})$ and $c(\mathbf{u}^{i}) < c(\mathbf{u}^{j})$. Then, we have
\begin{align}\label{eq:subcase1a}
a_{12}(\mathbf{u}^{i})& > a_{12}(\mathbf{u}^{j}) \Rightarrow \nonumber \\
a_{12}(\mathbf{u}^{i})f(p) + \sigma_{\mathbf{u}}^{2} &> a_{12}(\mathbf{u}^{j})f(p) + \sigma_{\mathbf{u}}^{2} \Rightarrow \nonumber \\
(1-\lambda) \frac{2 \sigma_{\mathbf{u}}^{2} f(p)}{ a_{12}(\mathbf{u}^{i})f(p) + \sigma_{\mathbf{u}}^{2}} &< (1-\lambda) \frac{2 \sigma_{\mathbf{u}}^{2} f(p)}{ a_{12}(\mathbf{u}^{j})f(p) + \sigma_{\mathbf{u}}^{2}},
\end{align}

\noindent
and
\begin{equation}\label{eq:subscase1b}
c(\mathbf{u}^{i}) < c(\mathbf{u}^{j}) \Rightarrow \lambda c(\mathbf{u}^{i}) < \lambda c(\mathbf{u}^{j}).
\end{equation}

\noindent
Combining (\ref{eq:subcase1a}) and (\ref{eq:subscase1b}), we get
\begin{align}
(1-\lambda) \frac{2 \sigma_{\mathbf{u}}^{2} f(p)}{ a_{12}(\mathbf{u}^{i})f(p) + \sigma_{\mathbf{u}}^{2}} + \lambda c(\mathbf{u}^{i}) < (1-\lambda) \frac{2 \sigma_{\mathbf{u}}^{2} f(p)}{ a_{12}(\mathbf{u}^{j})f(p) + \lambda c(\mathbf{u}^{j}) \sigma_{\mathbf{u}}^{2}} \Rightarrow \ell(p,\mathbf{u}^{i}) < \ell(p,\mathbf{u}^{j}),
\end{align}

\noindent
$\forall p \in [0,1]$. Since the last inequality holds for all $p \in [0,1]$, we conclude that ordering of controls can be achieved based on $a_{12}(\mathbf{u})$ independently of $p$.

\subsection{Proof of Corollary \ref{cor:CaseIV}}\label{ap:CaseIV}

We start from (\ref{eq:concave_current_cost_01}) and simplify terms as follows
\begin{align}
\ell(p,\mathbf{u}^{a}) \geqslant \ell(p,\mathbf{u}^{b}) &\Rightarrow \lambda c(\mathbf{u}^{a}) + (1-\lambda) \bigg ( 2f(p) - \frac{2 a_{12}(\mathbf{u}^{a}) f^{2}(p)}{a_{12}(\mathbf{u}^{a}) f(p) + \sigma_{1, \mathbf{u}^{a}}^{2}p + \sigma_{2, \mathbf{u}^{a}}^{2}(1-p)} \bigg ) 
\geqslant \lambda c(\mathbf{u}^{b}) \nonumber \\
&+ (1-\lambda) \bigg ( 2f(p) - \frac{2 a_{12}(\mathbf{u}^{b}) f^{2}(p)}{a_{12}(\mathbf{u}^{b}) f(p) + \sigma_{1, \mathbf{u}^{b}}^{2}p + \sigma_{2, \mathbf{u}^{b}}^{2}(1-p)} \bigg ) \Rightarrow \nonumber \\
& \frac{2 a_{12}(\mathbf{u}^{a}) f^{2}(p)}{a_{12}(\mathbf{u}^{a}) f(p) + \sigma_{1, \mathbf{u}^{a}}^{2}p + \sigma_{2, \mathbf{u}^{a}}^{2}(1-p)} \leqslant
\frac{2 a_{12}(\mathbf{u}^{b}) f^{2}(p)}{a_{12}(\mathbf{u}^{b}) f(p) + \sigma_{1, \mathbf{u}^{b}}^{2}p + \sigma_{2, \mathbf{u}^{b}}^{2}(1-p)} \Rightarrow \nonumber \\
&-2 a_{12}(\mathbf{u}^{a}) f^{2}(p) \big ( (\sigma_{2, \mathbf{u}^{a}}^{2} - \sigma_{2, \mathbf{u}^{b}}^{2}) + (\sigma_{1, \mathbf{u}^{a}}^{2} - \sigma_{1, \mathbf{u}^{b}}^{2} + \sigma_{2, \mathbf{u}^{b}}^{2} - \sigma_{2, \mathbf{u}^{a}}^{b})p \big ) \leqslant 0,
\end{align}

\noindent
where we have used that $a_{12}(\mathbf{u}^{a}) = a_{12}(\mathbf{u}^{b})$ and $c(\mathbf{u}^{a}) = c(\mathbf{u}^{b})$. We note that the term $-2 a_{12}(\mathbf{u}^{a}) f^{2}(p) \leqslant 0$. Therefore, the inequality is true if and only if
\begin{align}\label{eq:l1gl2}
(\sigma_{2, \mathbf{u}^{a}}^{2} - \sigma_{2, \mathbf{u}^{b}}^{2}) + (\sigma_{1, \mathbf{u}^{a}}^{2} - \sigma_{1, \mathbf{u}^{b}}^{2} + \sigma_{2, \mathbf{u}^{b}}^{2} - \sigma_{2, \mathbf{u}^{a}}^{2})p &~ \geqslant 0 ~ \Rightarrow \nonumber \\
& p ~ \geqslant ~ \frac{\sigma_{2, \mathbf{u}^{b}}^{2} - \sigma_{2, \mathbf{u}^{a}}^{2}}{\sigma_{1, \mathbf{u}^{a}}^{2} - \sigma_{1, \mathbf{u}^{b}}^{2} + \sigma_{2, \mathbf{u}^{b}}^{2} - \sigma_{2, \mathbf{u}^{a}}^{2}} \doteq p^{*},
\end{align}

\noindent
where we have exploited that $\sigma_{1,\mathbf{u}^{a}}^{2} > \sigma_{1,\mathbf{u}^{b}}^{2}$ and $\sigma_{2,\mathbf{u}^{a}}^{2} < \sigma_{2,\mathbf{u}^{b}}^{2}$. On the other hand, the inequality is false if and only if $p \leqslant p^{*}$.

\subsection{Proof of Lemma \ref{lem:SWWLB_formulae}}\label{ap:SWWLB_formulae_proof}

To determine the exact form of $G_{k+1,k+1}^{k+1}$, $G_{k+1,k}^{k+1}$, $G_{k,k+1}^{k+1}$ and $G_{k,k}^{k+1}$, we start from their definitions in Theorem 4.1 of \cite{RapoportTSP07}. First, we let
\begin{align}
L_{\ell}(\mathbf{y}_{\ell};x_{\ell}^{(1)},x_{\ell}^{(2)};x_{\ell-1};\mathbf{u}_{\ell - 1})& \doteq \frac{f(\mathbf{y}_{\ell} | x_{\ell - 1}^{(1)}, \mathbf{u}_{\ell - 1}) P(x_{\ell}^{(1)} | x_{\ell - 1}) }{f(\mathbf{y}_{\ell} | x_{\ell - 1}^{(2)}, \mathbf{u}_{\ell - 1}) P(x_{\ell}^{(2)} | x_{\ell - 1})}, \\
K_{\ell}(x_{\ell + 1};\mathbf{y}_{\ell};x_{\ell}^{(1)},x_{\ell}^{(2)};x_{\ell-1};\mathbf{u}_{\ell - 1})& \doteq \frac{P(x_{\ell + 1} | x_{\ell}^{(1)})}{P(x_{\ell + 1} | x_{\ell}^{(2)})} L_{\ell}(\mathbf{y}_{\ell};x_{\ell}^{(1)},x_{\ell}^{(2)};x_{\ell-1};\mathbf{u}_{\ell - 1}).
\end{align}

\noindent
Then, for the term $G_{k+1,k+1}^{k+1}$, we have that
\begin{align}
G_{k+1,k+1}^{k+1}& = \frac{\mathbb{E} \bigg \lbrace \bigg ( \sqrt{L_{k+1}^{+}(\mathbf{y}_{k+1})} - \sqrt{L_{k+1}^{-}(\mathbf{y}_{k+1})} \bigg )^{2} \bigg \rbrace}{\mathbb{E} \bigg \lbrace \sqrt{L_{k+1}^{+}(\mathbf{y}_{k+1})} \bigg \rbrace^{2}} = \frac{\mathbb{E} \big \lbrace L_{k+1}^{+}(\mathbf{y}_{k+1}) \big \rbrace - 2 \mathbb{E} \big \lbrace \sqrt{L_{k+1}^{+}(\mathbf{y}_{k+1}) L_{k+1}^{-}(\mathbf{y}_{k+1})} \big \rbrace}{\mathbb{E} \big \lbrace \sqrt{L_{k+1}^{+}(\mathbf{y}_{k+1})} \big \rbrace^{2}} \nonumber \\
&+ \frac{\mathbb{E} \big \lbrace L_{k+1}^{-}(\mathbf{y}_{k+1}) \big \rbrace}{\mathbb{E} \big \lbrace \sqrt{L_{k+1}^{+}(\mathbf{y}_{k+1})} \big \rbrace^{2}} \label{eq:Gkp1kp1kp1}
\end{align}

\noindent
where $L_{k+1}^{+}(\mathbf{y}_{k+1}) \doteq L_{k+1}(\mathbf{y}_{k+1};x_{k+1} + h_{k+1},x_{k+1};x_{k};\mathbf{u}_{k-1})$ and $L_{k+1}^{-}(\mathbf{y}_{k+1}) \doteq L_{k+1}(\mathbf{y}_{k+1};x_{k+1} - h_{k+1},\allowbreak x_{k+1};x_{k};\mathbf{u}_{k-1})$. We determine each term of (\ref{eq:Gkp1kp1kp1}) separately. Namely, we have
\begin{align}
\eta_{k}(h_{k+1},0)& \doteq \ln{\mathbb{E} \bigg \lbrace \sqrt{L_{k+1}^{+}(\mathbf{y}_{k+1})} \bigg \rbrace} = \ln\sum_{X^{k+1}} \int p(X^{k+1},U^{k},Y^{k+1})\frac{\sqrt{f(\mathbf{y}_{k+1}|x_{k+1} + h_{k+1},\mathbf{u}_{k})}}{\sqrt{f(\mathbf{y}_{k+1}|x_{k+1},\mathbf{u}_{k})}} \nonumber \\
& \times \frac{\sqrt{P(x_{k+1} + h_{k+1}|x_{k})}}{\sqrt{P(x_{k+1}|x_{k})}} d Y^{k+1} \overset{(a)}{=} \ln \sum_{x_{k}} P(x_{k}) \sum_{x_{k+1}} \sqrt{P(x_{k+1} + h_{k+1} | x_{k}) P(x_{k+1} | x_{k})} \nonumber \\
& \times \underbrace{\int \sqrt{f(\mathbf{y}_{k+1}|x_{k+1} + h_{k+1},\mathbf{u}_{k})} \sqrt{f(\mathbf{y}_{k+1}|x_{k+1},\mathbf{u}_{k})} d \mathbf{y}_{k+1}}_{\doteq \xi(x_{k+1} + h_{k+1}, x_{k+1})},\label{eq:Lkp1_plus_sqrt}
\end{align}

\noindent
where $(a)$ results from the Markovian nature of our system and the integral in (\ref{eq:Lkp1_plus_sqrt}) is the Bhattacharyya coefficient \cite{Duda01}
\begin{align}
\xi(x_{k+1} + h_{k+1}, x_{k+1})& = \int \sqrt{\mathcal{N}(\mathbf{m}_{x_{k+1}+h_{k+1}}^{\mathbf{u}_{k}}, \mathbf{Q}_{x_{k+1}+h_{k+1}}^{\mathbf{u}_{k}})} \sqrt{\mathcal{N}(\mathbf{m}_{x_{k+1}}^{\mathbf{u}_{k}}, \mathbf{Q}_{x_{k+1}}^{\mathbf{u}_{k}})} d \mathbf{y}_{k+1} \nonumber \\
&= \exp \bigg (  - \bigg [ \frac{1}{8} \big( \mathbf{m}_{x_{k+1}+h_{k+1}}^{\mathbf{u}_{k}} - \mathbf{m}_{x_{k+1}}^{\mathbf{u}_{k}} \big )^{T} \mathbf{Q}_{h}^{-1} \big( \mathbf{m}_{x_{k+1}+h_{k+1}}^{\mathbf{u}_{k}} - \mathbf{m}_{x_{k+1}}^{\mathbf{u}_{k}} \big ) \nonumber \\
&+ \frac{1}{2} \log \frac{\det \mathbf{Q}_{h} }{\sqrt{\det \mathbf{Q}_{x_{k+1}+h_{k+1}}^{\mathbf{u}_{k}} \cdot \det \mathbf{Q}_{x_{k+1}}^{\mathbf{u}_{k}}}} \bigg ]\bigg ),\label{eq:Bh_coefficient}
\end{align}

\noindent
Next, we have
\begin{align}
\ln{\mathbb{E} \bigg \lbrace L_{k+1}^{+}(\mathbf{y}_{k+1}) \bigg \rbrace}& = \ln\sum_{X^{k+1}} \int p(X^{k+1},U^{k},Y^{k+1})\frac{f(\mathbf{y}_{k+1}|x_{k+1} + h_{k+1},\mathbf{u}_{k}) P(x_{k+1} + h_{k+1}|x_{k})}{f(\mathbf{y}_{k+1}|x_{k+1},\mathbf{u}_{k}) P(x_{k+1}|x_{k})} d Y^{k+1} \nonumber \\
&=\ln \sum_{x_{k}} P(x_{k}) \sum_{x_{k+1}} P(x_{k+1} + h_{k+1} | x_{k}) \int f(\mathbf{y}_{k+1}|x_{k+1} + h_{k+1},\mathbf{u}_{k}) d \mathbf{y}_{k+1} = 0, \label{eq:Lkp1_plus}
\end{align}

\noindent
and similar is the case for $\ln{\mathbb{E} \big \lbrace L_{k+1}^{-}(\mathbf{y}_{k+1}) \big \rbrace}$. Finally, we have
\begin{align}
\ln{\mathbb{E} \bigg \lbrace \sqrt{L_{k+1}^{+}(\mathbf{y}_{k+1}) L_{k+1}^{-}(\mathbf{y}_{k+1})} \bigg \rbrace}& = \ln \sum_{x_{k}} P(x_{k}) \sum_{x_{k+1}} \sqrt{P(x_{k+1} + h_{k+1} | x_{k}) P(x_{k+1} - h_{k+1} | x_{k}) } \nonumber \\
& \times \underbrace{\int \sqrt{f(\mathbf{y}_{k+1}|x_{k+1} + h_{k+1},\mathbf{u}_{k})} \sqrt{f(\mathbf{y}_{k+1}|x_{k+1} - h_{k+1},\mathbf{u}_{k})} d \mathbf{y}_{k+1}}_{= \xi(x_{k+1} + h_{k+1}, x_{k+1}-h_{k+1})} \nonumber \\
&= \eta_{k}(h_{k+1},-h_{k+1}).\label{eq:Lkp1_plus_minus_sqrt}
\end{align}

\noindent
Substituting (\ref{eq:Lkp1_plus_sqrt}) --  (\ref{eq:Lkp1_plus_minus_sqrt}) back to (\ref{eq:Gkp1kp1kp1}) and exploiting the property $\exp (\ln ( \omega ) ) = \omega$, we get (\ref{eq:Gmatrix_recursive_00}).

Next, for the term $G_{k+1,k}^{k+1}$, we have
\begin{align}
G_{k+1,k}^{k+1}& = \frac{\mathbb{E} \big \lbrace \sqrt{L_{k+1}^{+}(\mathbf{y}_{k+1}) K_{k}^{+}(\mathbf{y}_{k})} \big \rbrace - \mathbb{E} \big \lbrace \sqrt{L_{k+1}^{+}(\mathbf{y}_{k+1}) K_{k}^{-}(\mathbf{y}_{k}) } \big \rbrace }{\mathbb{E} \big \lbrace \sqrt{L_{k+1}^{+}(\mathbf{y}_{k+1})} \big \rbrace \mathbb{E} \big \lbrace \sqrt{K_{k}^{+}(\mathbf{y}_{k})} \big \rbrace} \nonumber \\
& + \frac{-\mathbb{E} \big \lbrace \sqrt{L_{k+1}^{-}(\mathbf{y}_{k+1}) K_{k}^{+}(\mathbf{y}_{k}) } \big \rbrace + \mathbb{E} \big \lbrace \sqrt{L_{k+1}^{-}(\mathbf{y}_{k+1}) K_{k}^{-}(\mathbf{y}_{k}) } \big \rbrace }{\mathbb{E} \big \lbrace \sqrt{L_{k+1}^{+}(\mathbf{y}_{k+1})} \big \rbrace \mathbb{E} \big \lbrace \sqrt{K_{k}^{+}(\mathbf{y}_{k})} \big \rbrace}\label{eq:Gkp1kkp1}
\end{align}

\noindent
where $K_{k}^{+}(\mathbf{y}_{k}) \doteq K_{k}(x_{k+1};\mathbf{y}_{k};x_{k} + h_{k},x_{k};x_{k-1};\mathbf{u}_{k-1})$ and $K_{k}^{-}(\mathbf{y}_{k}) \doteq K_{k}(x_{k+1};\mathbf{y}_{k};x_{k} - h_{k},x_{k};x_{k-1};\mathbf{u}_{k-1})$. Next, we determine the four terms in the numerator and the term $\mathbb{E} \big \lbrace \sqrt{K_{k}^{+}(\mathbf{y}_{k})} \big \rbrace$ in the denominator. So, we have
\begin{align}
\rho_{k}(h_{k},0)& \doteq \ln{\mathbb{E} \bigg \lbrace \sqrt{K_{k}^{+}(\mathbf{y}_{k})} \bigg \rbrace} =  \sum_{x_{k-1}} P(x_{k-1}) \sum_{x_{k}} \sqrt{P(x_{k}|x_{k-1})} \sqrt{P(x_{k} + h_{k} | x_{k-1})} \nonumber \\ 
& \times \sum_{x_{k+1}} \sqrt{P(x_{k+1} | x_{k} + h_{k})} \sqrt{P(x_{k+1} | x_{k})} \underbrace{\int  \sqrt{f(\mathbf{y}_{k}|x_{k} + h_{k}, \mathbf{u}_{k-1})} \sqrt{f(\mathbf{y}_{k}|x_{k}, \mathbf{u}_{k-1})} d\mathbf{y}_{k}}_{= \xi(x_{k}+h_{k},x_{k})},\label{eq:Kk_plus_sqrt}
\end{align}

\noindent
and
\begin{align}
\zeta_{k}(h_{k},h_{k+1})& \doteq \ln{\mathbb{E} \bigg \lbrace \sqrt{L_{k+1}^{+}(\mathbf{y}_{k+1}) K_{k}^{+}(\mathbf{y}_{k})} \bigg \rbrace} = \ln \sum_{x_{k-1}} P(x_{k-1}) \sum_{x_{k}} \sqrt{P(x_{k} + h_{k} | x_{k-1}) P(x_{k} | x_{k-1})} \nonumber \\
&  = \sum_{x_{k+1}} \sqrt{P(x_{k+1} | x_{k} + h_{k})}\sqrt{P(x_{k+1} + h_{k+1} | x_{k})} \underbrace{\int \sqrt{f(\mathbf{y}_{k} | x_{k} + h_{k}, \mathbf{u}_{k-1})} \sqrt{f(\mathbf{y}_{k} | x_{k}, \mathbf{u}_{k-1})} d \mathbf{y}_{k}}_{\xi(x_{k}+h_{k},x_{k}}) \nonumber \\
&\times \underbrace{\int \sqrt{f(\mathbf{y}_{k+1} | x_{k+1} + h_{k+1}, \mathbf{u}_{k})} \sqrt{f(\mathbf{y}_{k+1} | x_{k+1}, \mathbf{u}_{k})} d \mathbf{y}_{k+1}}_{\xi(x_{k+1}+h_{k+1},x_{k+1})}.\label{eq:Lkp1_plus_Kk_plus}
\end{align}

\noindent
Similar to (\ref{eq:Lkp1_plus_Kk_plus}), for the rest denominator terms in (\ref{eq:Gkp1kkp1}), we get $\ln{\mathbb{E} \bigg \lbrace \sqrt{L_{k+1}^{+}(\mathbf{y}_{k+1}) K_{k}^{-}(\mathbf{y}_{k})} \bigg \rbrace} = \zeta_{k}(-h_{k},h_{k+1}), \allowbreak \ln{\mathbb{E} \bigg \lbrace \sqrt{L_{k+1}^{-}(\mathbf{y}_{k+1}) K_{k}^{+}(\mathbf{y}_{k})} \bigg \rbrace} = \zeta_{k}(h_{k},-h_{k+1}), \allowbreak \ln{\mathbb{E} \bigg \lbrace \sqrt{L_{k+1}^{-}(\mathbf{y}_{k+1}) K_{k}^{-}(\mathbf{y}_{k})} \bigg \rbrace} = \allowbreak \zeta_{k}(-h_{k},\allowbreak -h_{k+1})$. Substituting the above results back to (\ref{eq:Gkp1kkp1}) and exploiting the property $\exp (\ln ( \omega ) ) = \omega$, we get (\ref{eq:Gmatrix_recursive_01}). By symmetry, $G_{k+1,k}^{k+1} = G_{k,k+1}^{k+1}$.

Last but not least, for the term $G_{k,k}^{k+1}$, we have that
\begin{align}
G_{k,k}^{k+1}& = \frac{\mathbb{E} \bigg \lbrace \bigg ( \sqrt{K_{k}^{+}(\mathbf{y}_{k})} - \sqrt{K_{k}^{-}(\mathbf{y}_{k})} \bigg )^{2} \bigg \rbrace}{\mathbb{E} \bigg \lbrace \sqrt{K_{k}^{+}(\mathbf{y}_{k})} \bigg \rbrace^{2}} = \frac{\mathbb{E} \big \lbrace K_{k}^{+}(\mathbf{y}_{k}) \big \rbrace}{\mathbb{E} \big \lbrace \sqrt{K_{k}^{+}(\mathbf{y}_{k})} \big \rbrace^{2}} + \frac{- 2 \mathbb{E} \big \lbrace \sqrt{K_{k}^{+}(\mathbf{y}_{k}) K_{k}^{-}(\mathbf{y}_{k})} \big \rbrace + \mathbb{E} \big \lbrace K_{k}^{-}(\mathbf{y}_{k}) \big \rbrace}{\mathbb{E} \big \lbrace \sqrt{K_{k}^{+}(\mathbf{y}_{k})} \big \rbrace^{2}}.\label{eq:Gkkkp1}
\end{align}

\noindent
For the first term, we have
\begin{align}
\ln \mathbb{E} \big \lbrace K_{k}^{+}(\mathbf{y}_{k}) \big \rbrace = \ln \sum_{x_{k-1}} P(x_{k-1}) \sum_{x_{k}} P(x_{k} + h_{k} | x_{k-1}) \sum_{x_{k+1}} P(x_{k+1} | x_{k} + h_{k}) \int f(\mathbf{y}_{k} | x_{k} + h_{k}) d \mathbf{y}_{k} = 0.\label{eq:Kk_plus}
\end{align}

\noindent
Similarly, $\ln \mathbb{E} \big \lbrace K_{k}^{-}(\mathbf{y}_{k}) \big \rbrace = 0$. Finally, we have that
\begin{align}
\rho_{k}(h_{k},-h_{k})& \doteq \ln \mathbb{E} \bigg \lbrace \sqrt{K_{k}^{+}(\mathbf{y}_{k}) K_{k}^{-}(\mathbf{y}_{k})} \bigg \rbrace = \ln \sum_{x_{k-1}} P(x_{k-1}) \sum_{x_{k}} \sqrt{P(x_{k} + h_{k} | x_{k-1}) P(x_{k} - h_{k} | x_{k-1})} \nonumber \\
& \times \sum_{x_{k+1}} \sqrt{P(x_{k+1} | x_{k} + h_{k}) P(x_{k+1} | x_{k} - h_{k})} \underbrace{\int \sqrt{f(\mathbf{y}_{k} | x_{k} + h_{k}, \mathbf{u}_{k-1}) f(\mathbf{y}_{k} | x_{k} - h_{k}, \mathbf{u}_{k-1})} d \mathbf{y}_{k}}_{= \xi(x_{k} + h_{k},x_{k} - h_{k})}.\label{eq:Kk_plus_Kk_minus}
\end{align}

\noindent
Substituting (\ref{eq:Kk_plus_sqrt}), (\ref{eq:Kk_plus}) and (\ref{eq:Kk_plus_Kk_minus}) back to (\ref{eq:Gkkkp1}) and exploiting the property $\exp (\ln ( \omega ) ) = \omega$, we get (\ref{eq:Gmatrix_recursive_02}). 

For the information submatrix $J_{0}$, we have
\begin{align}
J_{0}& \doteq \frac{\mathbb{E} \bigg \lbrace \bigg ( \sqrt{L_{0}^{+}(\mathbf{y}_{0})} - \sqrt{L_{0}^{-}(\mathbf{y}_{0})} \bigg )^{2} \bigg \rbrace}{\mathbb{E} \bigg \lbrace \sqrt{L_{0}^{+}(\mathbf{y}_{0})} \bigg \rbrace^{2}} 
= \frac{\mathbb{E} \big \lbrace L_{0}^{+}(\mathbf{y}_{0}) \big \rbrace}{\mathbb{E} \big \lbrace \sqrt{L_{0}^{+}(\mathbf{y}_{0})} \big \rbrace^{2}} + \frac{- 2 \mathbb{E} \big \lbrace \sqrt{L_{0}^{+}(\mathbf{y}_{0}) L_{k+1}^{-}(\mathbf{y}_{0})} \big \rbrace + \mathbb{E} \big \lbrace L_{0}^{-}(\mathbf{y}_{0}) \big \rbrace}{\mathbb{E} \big \lbrace \sqrt{L_{0}^{+}(\mathbf{y}_{0})} \big \rbrace^{2}},\label{eq:J0_computation}
\end{align}

\noindent
where $L_{0}^{+}(\mathbf{y}_{0}) \doteq L_{0}(\mathbf{y}_{0};x_{0} + h_{0},x_{0};\mathbf{u}_{-1}) = \frac{p(\mathbf{y}_{0} | x_{0} + h_{0}, \mathbf{u}_{-1})}{p(\mathbf{y}_{0} | x_{0}, \mathbf{u}_{-1})} \allowbreak \times \frac{P(x_{0} + h_{0})}{P(x_{0})}$ and $L_{0}^{-}(\mathbf{y}_{0}) \doteq L_{0}(\mathbf{y}_{0};x_{0} - h_{0},x_{0};\mathbf{u}_{-1}) = \frac{P(x_{0} - h_{0})}{P(x_{0})} \allowbreak \times \frac{p(\mathbf{y}_{0} | x_{0} - h_{0}, \mathbf{u}_{-1})}{p(\mathbf{y}_{0} | x_{0}, \mathbf{u}_{-1})}$. First, we notice that
\begin{align}
\gamma(x_{0} + h_{0}, x_{0})& \doteq \ln{\mathbb{E} \bigg \lbrace \sqrt{L_{0}^{+}(\mathbf{y}_{0})} \bigg \rbrace} = \ln \sum_{x_{0}} \sqrt{P(x_{0} + h_{0}) P(x_{0})} \nonumber \\
& \times \underbrace{\int \sqrt{f(\mathbf{y}_{0} | x_{0} + h_{0}, \mathbf{u}_{-1})} \sqrt{f(\mathbf{y}_{0} | x_{0}, \mathbf{u}_{-1})} d \mathbf{y}_{0}}_{\xi(x_{0} + h_{0}, x_{0})},\label{eq:L0_plus_sqrt}
\end{align}

\noindent
$\ln{\mathbb{E} \big \lbrace L_{0}^{+}(\mathbf{y}_{0}) \big \rbrace} = \ln \sum_{x_{0}} P(x_{0} + h_{0}) \int f(\mathbf{y}_{0} | x_{0} + h_{0}, \mathbf{u}_{-1}) d \mathbf{y}_{0} = 0$ and $\ln{\mathbb{E} \big \lbrace L_{0}^{-}(\mathbf{y}_{0}) \big \rbrace} = \ln \sum_{x_{0}} P(x_{0} - h_{0}) \int f(\mathbf{y}_{0} | x_{0} - h_{0}, \mathbf{u}_{-1}) d \mathbf{y}_{0} = 0$. Next, we have that $\gamma(x_{0} + h_{0}, x_{0} - h_{0})$ is
\begin{align}
\ln{\mathbb{E} \bigg \lbrace \sqrt{L_{0}^{+}(\mathbf{y}_{0}) L_{0}^{-}(\mathbf{y}_{0})} \bigg \rbrace}& = \ln \sum_{x_{0}} \sqrt{P(x_{0} + h_{0}) P(x_{0} - h_{0})} \nonumber \\
& \times \underbrace{\int \sqrt{f(\mathbf{y}_{0} | x_{0} + h_{0}, \mathbf{u}_{-1})} \sqrt{f(\mathbf{y}_{0} | x_{0} - h_{0}, \mathbf{u}_{-1})} d \mathbf{y}_{0}}_{\xi(x_{0} + h_{0}, x_{0} - h_{0})}\label{eq:L0_plus_sqrt_L0_minus_sqrt}
\end{align}

\noindent
Substituting (\ref{eq:L0_plus_sqrt})--(\ref{eq:L0_plus_sqrt_L0_minus_sqrt}) back to (\ref{eq:J0_computation}) and exploiting the property $\exp (\ln ( \omega ) ) = \omega$, we determine the final form of $J_{0}$.

\end{appendix}

\bibliographystyle{IEEEtran}
\bibliography{IEEEabrv,references}


\end{document}